\documentclass[10pt]{article}

\pdfoutput=1
\usepackage[round, sort, numbers, authoryear]{natbib} 
\usepackage[utf8]{inputenc}
\usepackage[L7x]{fontenc}
\usepackage{multirow}
\usepackage[american]{babel} 
\usepackage{lmodern}
\usepackage{longtable}
\usepackage{enumerate}
\usepackage{pgfplotstable}
\usepackage{graphicx}
\usepackage{epstopdf}
\usepackage{a4wide}
\usepackage[all]{xy}
\usepackage{array}
\usepackage{tabularx}
\usepackage{booktabs}
\usepackage{colortbl}
\usepackage{verbatim}
\usepackage{rotating}
\usepackage{xtab}
\usepackage{lmodern}
\usepackage[top=2cm, bottom=2cm, left=3cm, right=3cm, footskip=1cm, a4paper]{geometry}
\usepackage{amssymb, amsmath, amsthm}
\usepackage{etoolbox}
\usepackage{morefloats}
\usepackage{footnote}
\usepackage[toc,page]{appendix}
\usepackage[bottom, splitrule]{footmisc}
\usepackage{float}
\usepackage[font=small]{caption}
\usepackage{chngcntr}
\usepackage{dsfont}
\usepackage{tipa}
\usepackage{covington}
\usepackage[nottoc]{tocbibind}
\usepackage{enumitem}
\usepackage{hyperref}
\newenvironment{sciabstract}{%
	\begin{quote} \small}
	{\end{quote}}

\newcounter{lastnote}

\title{
	\vskip 15pt
	Business Cycle Synchronization in the EU: A Regional-Sectoral Look through Soft-Clustering and Wavelet Decomposition
	}

\author
{Saulius Jokubaitis,$^{1, 2\ast}$ Dmitrij Celov$^{1, 2}$\\
	\\
	\normalsize{$^{1}$Faculty of Mathematics and Informatics, Vilnius University, Lithuania}\\
	\normalsize{$^{2}$Faculty of Economics and Business Administration, Vilnius University, Lithuania}\\
	\\
	\normalsize{$^\ast$ To whom correspondence should be addressed; E-mail:  saulius.jokubaitis@mif.vu.lt}
}

\date{}

\addto{\captionsenglish}{}
\addto{\captionsbritish}{}
\addto{\captionsamerican}{}
 
\begin{document} 
	
	\maketitle 

	\begin{sciabstract}
	This paper elaborates on the sectoral-regional view of the business cycle synchronization in the EU -- a necessary condition for the optimal currency area. We argue that complete and tidy clustering of the data improves the decision maker's understanding of the business cycle and, by extension, the quality of economic decisions. We define the business cycles by applying a wavelet approach to drift-adjusted gross value added data spanning over 2000Q1 to 2021Q2. For the application of the synchronization analysis, we propose the novel soft-clustering approach, which adjusts hierarchical clustering in several aspects. First, the method relies on synchronicity dissimilarity measures, noting that, for time series data, the feature space is the set of all points in time. Then, the ``soft'' part of the approach strengthens the synchronization signal by using silhouette measures. Finally, we add a probabilistic sparsity algorithm to drop out the most asynchronous ``noisy'' data improving the silhouette scores of the most and less synchronous groups. The method, hence, splits the sectoral-regional data into three groups:  the synchronous group that shapes the EU business cycle; the less synchronous group that may hint at cycle forecasting relevant information; the asynchronous group that may help investors to diversify through-the-cycle risks of the investment portfolios. Our results do not contradict the core-periphery hypothesis, suggesting that France, Germany, Austria and Italy countries and more export-oriented economic activities such as an industry, trade, transportation, and tourism, drive the EU business cycle. Catching-up small open Estonia's economy, in this context, is closely following the synchronous cycle and may act as a barometer of the EU cycle. The less synchronous group consists of agriculture, public services, and financial services that respond to global shocks to a lesser extent. Finally, the drop-out segment includes periphery regions, agriculture and more closed-sector-oriented services.

		
		
	\end{sciabstract}
	
	\section*{Introduction}
	
	Since the introduction and later adoption of the euro as a common European currency in 1999, both academia and market participants have been interested in the phenomenon of business cycles' synchronization among European economies. The outbreak of the global financial crisis in 2008 and the consequent European sovereign debt crisis revived debates about the optimality of the European Monetary Union (EMU) due to observed discrepancies between groups of countries within and outside the euro area. In this context, the most lingering question underlying the Optimal Currency Area (OCA) theory is understanding if the euro area members are similar enough to share the same currency in the long run. If they are, does the homogeneity lead to a core-periphery division, and of what complexity (\cite{bengoechea2006useful}). The latter question can be addressed through a grouping or clustering perspective, considering the cyclical (dis-)similarities between the core and more `vulnerable' regions (see, among others, \cite{fidrmuc2006meta}, \cite{stanivsic2013convergence}, \cite{di2016business}, \cite{ahlborn2018core}, \cite{rathke2022similar}, \cite{arcabic2022business}, \cite{de2022coherence}). 
	
	Mitigating the risk of the euro area fragmentation, in 2012, the European institutions undertook unprecedented actions. The ECB, responsible for the EMU monetary policy, contributed by using the so-called `quantitative easing' strategy, while the public sector of the core economies -- by applying for purchase programs. At the same time, the periphery countries significantly tightened their fiscal policy and banking sector supervision  (\cite{COUDERT2020428}). The importance of business cycle synchronization comes then into play as the necessary OCA condition. Otherwise, for the monetary union with asynchronous regional business cycles, the single monetary policy will not optimally stabilize the common shocks for some of its members.
    Besides, if the shapes of synchronized cycles are different, individual policy responses to global economic shocks may be too accommodative for some countries, and too tight for the others. This mismatch implies the smooth state changes extending the duration of recoveries from undesirable overheating-recession states. Furthermore, the economic policies may last too long for countries with shorter business cycles and too fast for countries with lengthier ones (\cite{bengoechea2006useful}). 
	
	Examination of business cycles and their components goes back to the OCA theory (see, among others,  \cite{mckinnon1963optimum}, \cite{frankel1998endogenity}). In these papers, the authors argued that the lack of an independent monetary policy potentially leads to a significant loss of welfare and even a breakdown of a monetary union if the members exhibit asymmetric or asynchronous output fluctuations. On the one hand, \cite{krugman1991increasing} argues that increasing integration might lead to regional concentration of industrial activities, which would, in turn, result in sectoral-regional specific shocks, thereby increasing the likelihood of asymmetric shock and diverging business cycles. Furthermore, economic integration through more intense capital flows might lead to a more diverse production structure of the union member states and a trade growth, therefore, improving the business cycle synchronization (\cite{kalemli2001economic}). On the other hand, \cite{frankel1998endogenity} suggests that the removal of trade barriers will lead to easier transmission of any demand shocks across countries, implying more symmetric fluctuations in the business cycles. \cite{inklaar2008trade} add that as the economic policies in the euro area become more aligned, the business cycle synchronization should also improve. 
	
	In this regard, the purpose of current paper is to revisit the OCA business cycle synchronization controversy by analysing the sectoral-regional view of EU-27 Gross Value Added (GVA) data. In other words, we seek to elaborate on the regional dimension by scrutinising its sectoral split as the clusters of economic activities, which may hint at two things. First, the policymakers might choose the relevant economic policies, mitigating the negative overheating or downturn impacts of the synchronous European regions and sectors, and predict the extent of contagion effects stemming from various economic shocks. Second, the less synchronous and asynchronous EU member states and economic activities hint to investors which areas are potentially interesting for portfolio diversification as their cyclical components move against the joint European economic cycle. Therefore, the methods described in the paper equip researchers and decision-makers with new data that helps understanding the sectoral-regional composition of (a)-synchronous business cycle groups in Europe.
	
	From methodological perspective, we contribute to the business cycle synchronization literature in several ways. First, we introduce the relevant methodological framework that combines wavelet decomposition with soft-clustering techniques for identifying the growth cycles. As the single euro area monetary policy directly mitigates demand-driven shocks, we focus on European business cycles observed within 2-8 years window. Second, to improve the clustering results and demonstrate the evidence of synchronization, we propose a synchronization-based dissimilarity measure for clustering. Third, we use several cluster cleaning approaches based on silhouette scores and the bootstrapped soft-clustering probabilities allowing us to control the concentration and tightness of the estimated groups in synchronization. We find that the optimal number of clusters is small. For the purpose of the paper it is sufficient to study three groups -- two purified clusters and the least synchronous drop-out group, which supports the core-periphery literature in the context of the regional-sectoral breakdown of the European economic activities. All these proposals eventually lead to an efficient approach for obtaining the clean synchronous and asynchronous clusters of EU regions and economic activities that fit the different needs of policymakers and investors.
	
The paper is structured as follows. In Section 1, we present the methodology introducing the soft-clustering approach. In Section 2, we discuss the features of the data, the trade-offs when choosing the set of hyperparameters, and elaborate on the main results of applying the soft-clustering approach to EU-27 GVA data for 11 segments of the economy. In this section, we primarily discuss synchronous cycle composition and elaborate on the core-periphery hypothesis. Section 3 provides concluding remarks.

	\section{Methodology}

    \subsection{Data}

	In this paper, we scrutinize quarterly, seasonally and calendar-adjusted, real (in chain-linked volumes sense, index 2010=100) GVA data sourced from Eurostat,  spanning over $\mathcal{T} = \{2000Q1, \ldots, 2021Q2\}$, $|\mathcal{T}| = 86$ quarters. Seeking to analyse the regional-sectoral view, the dataset contains 27 EU countries (after Brexit, excluding the UK) and 11 sectors that correspond to A*10 economic activities breakdown taken from NACE rev.~2. The latter breakdown broadly defines the industry as B-E economic activities, excluding construction (F). The sectoral breakdown also narrowly defines the industry as manufacturing activities (C). Since the inclusion of C allows choosing more relevant economic policies or building more diversified investment portfolios, we keep both industry definitions. For simplicity, denoting the EU country indices as $S = \{ 1, \ldots, 27\}$ and the different economic activities as $R = \{1, \ldots, 11\}$, we construct the set of indices $\mathcal{I}:= R \times S$, where $\times$ denotes the Cartesian product and $|\mathcal{I}| = 297$. In sum, throughout the paper, we analyse 297 time series, denoting any such series as $X_{i}(t)$, where the indices $ i \in \mathcal{I}$ correspond to a specific regional-sectoral indicator, measured at a time point $t \in \mathcal{T}$. In this context, $\mathcal{T}$ represents the set of 86 features recorded for each time series.
	
	\subsection{Wavelet decomposition}\label{sec:wavelet}
	By definition, any trend-cycle decomposition method decomposes macroeconomic data into low-frequency trend $X^*_{i}(t)$ that may have both stochastic and deterministic parts, a cycle $c_{i}(t)$, likely consisting of sub-cycles of different lengths, and high-frequency irregular noise $\nu_{i}(t)$, typically interpreted as shocks (\cite{celovcomunale2021}):
	\begin{eqnarray} \label{eq:TCD}
		X_{i}(t) &:=& X^*_{i}(t) + c_{i}(t) + \nu_{i}(t). 
	\end{eqnarray}
	
	The data preparation step then has to isolate the growth cycle by removing the trend and noise with the least possible distortions. For this purpose, we employ the wavelet decomposition approach.
	
	The wavelet transformation is a powerful signal decomposition tool that uses both the time and frequency domains. For applications in the business cycle literature, see, for example, \cite{crowley2009fused}, \cite{bruzda2011business}, \cite{soares2011business},  \cite{AGUIARCONRARIA2011477}, among others. 
	
	Following \cite{RePEc:eee:finlet:v:19:y:2016:i:c:p:298-304}, we employ the maximal overlap discrete wavelet transformation (MODWT) approach with $J=5$. We  apply the Least Asymmetric wavelets with 8 parameters (LA(8)) for the smoother results. Using lengthy wavelets may induce smoothing artefacts at the ends-of-sample, hence, require cautious interpretation of the end-points. Although the MODWT transformation is not orthonormal and is highly redundant, it admits any sample sizes $\mathcal{T}$ that need not be the multipliers of a power of two. Besides, the chosen transformation is invariant to circular shifts and can decompose the data by applying multiresolution analysis. For the latter, using the efficient pyramid algorithm under any base functions (e.g., Haar, Least Asymmetric, or Fejér-Korovkin), the data decomposes to:
		\begin{eqnarray} \label{eq:decomp}
		X_{i}(t) &:=& \sum^{J}_{j=1} D_{j}^{(i)}(t) + S_{J}^{(i)}(t),~~i \in \mathcal{I},~ t \in \mathcal{T}, 
	\end{eqnarray}
	where $D_{j}^{(i)}(t)$ is the $j$-th level wavelet and $S_{J}^{(i)}(t)$ is the smooth component.
	
	\cite{crowley2009fused} argue that, for business cycle frequencies, it is sufficient to analyse $D_3$ and $D_4$ wavelets, spanning over 2-8 year cycle lengths (see Table \ref{table:wavelet1}). The choice may be viewed as a combination of cyclical fluctuations stemming from variation in inventories (Kitchin) and fixed investments (Juglar) correspondingly. However, we prefer to interpret the business cycle as an aggregate outcome of supply-demand interplay from some global dynamic general equilibrium model that reflects sectoral-regional split of the data and demonstrates responses to a single euro area monetary policy designed to mitigate demand-driven shocks. If one would be interested in studying macro-prudential and fiscal policy impacts, we would also recommend including the credit cycle reflecting component $D_5$ into the common cycle. \cite{celovcomunale2021} argue that omission of this component impacts mostly the amplitude of the waves as the business cycle wavelets are placed on top of smoother and longer credit cycle wave. Hence, the local turning points of the common cycle reflect more business cycle characteristics than that of the smoother $D_5$ wavelet.
	
	\begin{table}[ht!]
				\caption{Frequency interpretation for quarterly data \cite{crowley2009fused}} \label{table:wavelet1}
		\centering
		\begin{tabular}{c|c|c}
			Detail&Length&Type\\ \hline
			$D_1$&2--4Q&Noise\\
			$D_2$&1--2Y&Noise\\
			$D_3$&2--4Y&Business Cycle\\
			$D_4$&4--8Y&Business Cycle\\
			$D_5$&8--16Y&Credit Cycle\\
			$S_5$&16Y+&Trend\\
		\end{tabular}

	\end{table}

	Using LA(8) base functions, we combine the 297 extracted wavelet series $D_{3}^{(i)}(t)$ and $D_{4}^{(i)}(t)$ from the variables $X_{i}(t)$, $i \in \mathcal{I}, t \in \mathcal{T}$, defining the business cycle frequencies related growth cycle $c_{i}(t)$ by: 
	\begin{eqnarray} \label{eq:cycle}
		c_{i}(t) &:=& D_{3}^{(i)}(t) + D_{4}^{(i)}(t),~~i \in \mathcal{I},~ t \in \mathcal{T}. 
	\end{eqnarray}

	The form of the cycle $c_{i}(t)$ returned by wavelet decomposition is sensitive to the appropriate trend elimination procedure applied to processing the data for MODWT. First, following \cite{celovcomunale2021}, we use the drift-adjusted time series data:
		\begin{eqnarray} \label{eq:drift_adj}
		\tilde{X}_{i}(t) &:=& X_{i}(t) - \frac{(t-1)(X_{i}(T) - X_{i}(1))}{T-1}. 
	\end{eqnarray}

Note, that for notational convenience throughout this paper we will assume that $X_{i}(t)$ denotes the drift-adjusted series by \eqref{eq:drift_adj}. Second, since the MODWT filter is circular, the way the edges indefinitely attach may significantly distort the end-of-sample behaviour of the estimated cycles. We use the ``reflective'' approach, which extends the observed data by indefinite mirroring of the data. Unlike the ``periodic'' extension, a reflection of the latest data should, in theory, keep the signals similar at the ends, while the former may experience spurious structural breaks. ``Reflective'' connection roughly assumes that the static synchronicity signal goes through possibly unobserved data. However, macroeconomic data may experience time-varying synchronicity between different cycles changing after some global or local events, for instance, due to the financial crisis, COVID-19 outbreak, or after adopting the euro by new euro area members. The drift-adjustment \eqref{eq:drift_adj} makes the distortions at the ends-of-sample less apparent.
	
	In this paper, the remainder series $D_5^{(i)}(t) + S_5^{(i)}(t)$ determine the stochastic part of the trend of the original drift-adjusted data that also contains credit cycle associated wavelet, while the first two wavelets define the noise. 
	We observe that the applied definition of a trend is sufficient for most of the analysed GVA data. The trends are similar to those obtained with the Hodrick-Prescott (HP) filter with penalty $\lambda$ optimised for the 8-years cycles. The HP filter is widely used in the applications (see, e.g., \cite{dickerson1998business},  \cite{inklaar2008trade}, \cite{papageorgiou2010business}), the robustness of which is confirmed by \cite{artis1997international} and \cite{dickerson1998business}, among others. Hence, the HP filter is an appropriate benchmark for a robustness check. We observe the median correlation of $0.996$ between the series (smallest -- 0.98), with scaled and centred RMSE between the two extracted trends falling within $(0.01, 0.32)$, 90\% of which is lower than 0.15. The deviation is more visible at the ends-of-sample, increasing only marginally to 0.2, where both approaches experience larger distortions.
	Following these observations, we deem the trend elimination
	using wavelets sufficient in our case. 
	
	\subsection{Synchronicity measures}
	For the analysis of synchronization, we use the synchronicity measure as in \cite{mink2012measuring}, defined below by \eqref{eq:varphi_1}--\eqref{eq:varphi_2}. For every time period considered, a value of 1 indicates that the two cycles have the same sign, while a value of -1 denotes cycles of different signs. In other words, for any time period $t \in \mathcal{T}$ and $i, j \in \mathcal{I},\ \ i \neq j$, the synchronicity measure is:
	\begin{eqnarray} \label{eq:varphi_1}
		\varphi(i,j)(t) &:=& \frac{c_{i}(t) c_{j}(t)}{|c_{i}(t) c_{j}(t)|} \in \{-1, 1\}, 
	\end{eqnarray}
	where, by \eqref{eq:cycle}, $c_{i}(t)$ denotes the cycle of indicator $i$ at time moment $t$. Averaging over the total feature space $\mathcal{T}$ defines the through-the-cycle synchronicity measure: 
	\begin{eqnarray}\label{eq:varphi_2}
		\varphi^{(\mathcal{T})}(i,j) := \frac{1}{|\mathcal{T}|}\sum_{t \in \mathcal{T}} \varphi(i,j)(t), 
	\end{eqnarray}
where $\varphi^{(\mathcal{T})}(i,j) \in [-1,1]$ for any pair of indicators $i \neq j \in \mathcal{I}$. The closer the  $\varphi^{(\mathcal{T})}(i,j)$ estimates are \mbox{to 1}, the stronger are the co-movements between the indicator pair $(i,j)$ over the period $ \mathcal{T} $.

 	In addition, to validate the quality of business cycles synchronization for the group of indicators, we consider the measure proposed by  \cite{RePEc:eee:jmacro:v:37:y:2013:i:c:p:265-284} and defined in the indicator space $\mathcal{I}$ as follows:
		\begin{eqnarray}\label{eq:js1}
		\gamma^{\mathcal{I}}(i,t) &:=& \log(s_{-i}^{\mathcal{I}}(t)) - \log(s^{\mathcal{I}}(t)),
	\end{eqnarray}
where $s_{-i}^{\mathcal{I}}(t)$ denotes the standard deviation within the group $\mathcal{I}$, excluding the indicator $i$, at a fixed time point $t\in\mathcal{T}$, and $s^{\mathcal{I}}(t)$ is the standard deviation including the indicator $i$:
\begin{eqnarray*} 
	(s_{-i}^{\mathcal{I}}(t))^2 &:=& \sum_{\substack{j \in \mathcal{I} \\ j\neq i}}\bigg(c_{j}(t) - \frac{1}{|\mathcal{I}|-1}\sum_{\substack{j' \in \mathcal{I} \\ j'\neq i}} c_{j'}(t) \bigg)^2, \\
	(s^{\mathcal{I}}(t))^2 &:=& \sum_{j \in \mathcal{I}}\bigg(c_{j}(t) - \frac{1}{|\mathcal{I}|}\sum_{j' \in \mathcal{I}} c_{j'}(t) \bigg)^2,
\end{eqnarray*}
for some $i \in \mathcal{I}, |\mathcal{I}|>1$.  
The main idea is that the exclusion of any indicator $i$ from some group $\mathcal{I}$ of countries (e.g., euro area or the EU) or sectoral-regional indicators, as in our case, can either increase or decrease the total within-group variance. If we assume that the signals are synchronized and concentrated around the mean cycle, we will observe consistently small standard deviations through $t \in \mathcal{T}$. Thus, the removal of any highly-concentrated indicator $i$ from the group $\mathcal{I}$ should only increase the total variance of the group. On the other hand, if certain signals are far away from the mean consistently, removal of those from the group will decrease the variance, improving the synchronization of the remaining members of the club.
Following these ideas, we compute the median statistic over $t \in \mathcal{T}$:
\begin{eqnarray} \label{eq:js2}
	\gamma^{\mathcal{I}}_{i} &:=& \operatorname{Med}_{t \in \mathcal{T}}\left(\gamma^{\mathcal{I}}(i,t)\right).
\end{eqnarray} 
We argue that \eqref{eq:js2} is the robust proximity of the synchronization measure. A certain degree of robustness is achieved by taking the median over $t \in \mathcal{T}$, countering against periods with high volatility that may introduce large leverage on the whole series.  A negative sign of \eqref{eq:js2} suggests that the synchronization improves upon excluding indicator $i$ from the group $\mathcal{I}$, while the positive sign shows that the situation is worse without indicator $i$. In addition, this measure captures the proximity of an indicator $i$ to the mean cycle within the group. Note, that if all indicators within any group are in perfect synchronization (e.g., resulting in $\varphi^{(\mathcal{T})}(i,j) =1$, $i\neq j$), the cycles can still vary in their amplitude. Thus, with sufficiently high variation, the \eqref{eq:js2} measure will always find a subset of indicators with negative signs of \eqref{eq:js2}, unless perfect co-variation occurs. Nevertheless, the measure can be useful to gauge how close are the signals to the mean signal within the group/cluster $\mathcal{I}$. Finally, the underlying idea behind \eqref{eq:js1} and \eqref{eq:js2}, that certain elements of the cluster can reduce the overall synchronicity of the cluster, gave rise to the probabilistic soft-clustering approach, introduced in Section \ref{sec:soft}.
	
	\subsection{Clustering}
	In the literature, there are many examples of clustering methods applied in the context of macroeconomic time series or related to the business cycle synchronization (see, e.g., \cite{artis2002membership}, \cite{papageorgiou2010business}, \cite{ahlborn2018core}, \cite{COUDERT2020428}). In this paper, we consider hierarchical clustering (see, \cite{COUDERT2020428} and references within) as the base clustering approach with Ward's minimum variance linkage method that aims at finding compact, spherical clusters (\cite{ward2}). The main novelty of the paper is the application of  synchronicity $\varphi^{(\mathcal{T})}(i,j)$ as the dissimilarity measure together with a soft-clustering approach -- the adjusted version of hierarchical clustering algorithm combined with silhouette scores (see Section \ref{sec:sil}) and bootstrapped probabilistic thresholding (see Section \ref{sec:soft}). For hierarchical clustering, by \eqref{eq:varphi_2}, we construct a dissimilarity measure \begin{eqnarray}
		d^{(\mathcal{T})}_{i,j} := 1 - \varphi^{(\mathcal{T})}(i,j),
	\end{eqnarray}
	where the closer the $\varphi^{(\mathcal{T})}(i,j)$ is to 1, the less dissimilar is the specific pair of indicators $(i,j)$. 
	Throughout the paper, we define the dissimilarity matrix as:
	\begin{eqnarray}\label{eq:dist}
		D^{(\mathcal{T})} := (d^{(\mathcal{T})}_{i,j})_{i,j=1}^{n}.
	\end{eqnarray}
	
	The use of the whole period $\mathcal{T}$ for $\varphi^{(\mathcal{T})}(i,j)$ allows us to gauge the overall synchronization levels between the considered cycles. However, such approach can miss certain vanishing or very recent co-movements due to averaging. Furthermore, the approach is useful to establish strong co-movements between the cycles' pair $(i,j)$ but is less informative for weaker co-movements. Note, that \eqref{eq:varphi_1} and \eqref{eq:varphi_2} essentially counts the cycle movements of the same sign throughout the set $\mathcal{T}$ but is not affected by the ordering of the series. In other words, the measure does not reward sequential co-movements of the cycles, losing part of the information from the data. To account for these aspects, in the soft-clustering step, we use dissimilarities $D^{(\mathcal{T})}$ along with a established thresholding rule that enables us to re-weight the cluster probabilities based on the observed synchronization strength (see Section \ref{sec:soft}). The thresholding set of hyperparameters allows us to consider only those indicators that show strong synchronization throughout the larger part of the period $\mathcal{T}$, dismissing the weaker co-moving signals. In fact, in Section \ref{sec:eu_comp} we demonstrate that the excluded cycles fall into the group of asynchronous cycles,
	the dynamics
	of which can potentially be very useful, e.g., for investment risk diversification.
	
	Furthermore, seeking to reduce the number of dismissed signals, we could split the time horizon $\mathcal{T}$ into a certain number $s \geq 1$ of smaller time windows $\mathcal{T} = \cup_{j=1}^{s} \mathcal{T}_{j}$ and find the corresponding dissimilarity matrices $D^{(\mathcal{T}_{j})}$, $j = 1, \ldots, s$. In this case, the clustering procedure can be applied for every time window $\mathcal{T}_{j}$. Combining all the resulting clusters through a soft-clustering approach would then introduce a certain reward for those cycles, that show high synchronicity during specified time windows $\mathcal{T}_{j}$. In this paper we consider such a procedure with $s = 3$ time windows, namely, $\mathcal{T}_{1} = \{2000Q1, \ldots, 2007Q4\}$, $\mathcal{T}_{2} = \{2008Q1, \ldots, 2014Q4\}$ and $\mathcal{T}_{3} = \{2015Q1, \ldots, 2021Q2\}$. The idea behind the choice of the splitting dates is to separate the co-movements after introduction of the euro, the shock of the 2008 financial crisis and followed sovereign debt crisis (W-recovery), and the period that ends with COVID-19 outbreak. Similar approach was taken, for instance, by \cite{COUDERT2020428}, \cite{bunyan2020fiscal}. 
	
	\subsection{Silhouette scores}
	\label{sec:sil}
	Since clustering is an unsupervised learning approach, it is impossible to use cross-validation. Instead, we propose a soft approach in order to improve the homogeneity of the synchronous cycles containing clusters. For this purpose, we employ silhouette scores, which show promising results and consistently good performance in various applications and simulations (see, e.g., \cite{arbelaitz2013extensive}). High observed silhouette score values would effectively hint on the level of homogeneity in the synchronous cluster of business cycles.
	
	The general algorithm follows \cite{ROUSSEEUW198753}, \cite{artis2002membership}. Given a cluster $C_{m}$, some indicator $i \in C_{m}$, and a dissimilarity function $d_{i,j}$, define 
	\begin{eqnarray}
		a_i &:=& \frac{1}{|C_m|-1}\sum_{\substack{j \in C_{m} \\ j \neq i}} d_{i,j}.
	\end{eqnarray}
	Next, for any $ C_{k}, m\neq k$, define
	\begin{eqnarray}
		b_i &:=& \min_{k: ~k \neq m} \frac{1}{|C_{k}|}\sum_{j \in C_{k}}d_{i,j}. 
	\end{eqnarray}
	Any cluster $C_{k}$, corresponding to the smallest value $b_i$, determines the neighbouring cluster for the indicator $i$. Thus, for any $i$, the silhouette score $s_i$ is defined as:
	\begin{eqnarray}
		s_i&:=& \frac{b_i - a_i}{\max\{a_i, b_i\}} \in [-1,1], ~~\text{ if }|C_{m}|>1,
	\end{eqnarray}
	and $s_i=0$ if $|C_{m}|=1$. If, for any indicator $i$, the resulting silhouette score $s_i < 0$, the method suggests that there exists a neighbouring cluster, where the indicator $i$ fits better. 
	
	Throughout this paper,  we employ the silhouette scores in two ways. First, we use the scores for the choice of the number of clusters when performing hierarchical clustering. For any resulting cluster $C_i$, we estimate the average silhouette score within the cluster, i.e., calculate
	\begin{eqnarray}
		SC_i &:=& \frac{1}{|C_i|}\sum_{j \in C_{i}}s_j.
	\end{eqnarray}
	For a given number of clusters $k > 1$, we estimate $SC_{i}, ~i = 1,\ldots,k$. The aim is to choose such values of $k$, for which both $\min_{1 \leq j \leq k}SC_{j}$ and $k^{-1}\sum_{j=1}^{k}SC_j$ are the largest.

	Second, we use the scores $s_i$ to clean the resulting cluster $C_i$ retaining only the indicators with scores above a specified threshold $\kappa \in \mathbb{R}$. The steps for tidying the data are outlined below:
	
	\begin{enumerate}
		\item For every $i = 1, \ldots, k$, consider a cluster $C_i$.
		\begin{enumerate}
			\item For every $j \in C_i$, estimate the corresponding Silhouette scores $s_j$.
		\end{enumerate}
		\item Define $\tilde C_i := C_i$, $i = 1, \ldots, k$.
		\item Collect all scores from step (1.). If all $s_j > \kappa$, stop the algorithm, else, proceed to (4.). 
		\item Using the scores from step (1.), remove every indicator $j \in \tilde C_i$, if $s_j \leq \kappa$. 
		\item Repeat step (1.) with the updated clusters $\tilde C_j$. 
	\end{enumerate}
Throughout this paper we assume that $\kappa = 0$. Therefore, the algorithm stops only when all of the corresponding silhouette scores are positive. However, the downside of such approach is that it can discard too many variables of interest from the analysis due to step (4.) of the algorithm. As a general case, one should consider a range of values for $\kappa$ and fine-tune based on the expected percentage of variables discarded. In our case, setting $\kappa = 0$ was sufficient and resulted in relatively small number of discarded variables from the analysis. Note, that excluded variables do not contribute further for the establishment of synchronizing clusters, however are still valid members of the pool consisting of drop-out indicators, likely capturing the asynchronous growth cycles, which may be of further interest for different investment diversification strategies or economic development policies.

	\subsection{A soft-clustering approach}
	\label{sec:soft}
	As the main result of the paper, we propose the following soft-clustering algorithm, which requires setting hyperparameters $(\omega_1, \omega_2, \ldots, \omega_6)$: 
	
	\begin{enumerate}
		\item For iteration $b = 1, \ldots, \omega_1$, where $\omega_1$ denotes the number of bootstrapped samples, repeat:
		\begin{enumerate}
			\item Draw a random subset $\{X_{J_b}(t),~ t \in \mathcal{T} \}$, where $J_b \subset \mathcal{I}$ is a bootstrapped sample of indices $\mathcal{I}$ of the same size as $|\mathcal{I}|$ drawn with replacement. 
			On average, 63.3\% of indices are unique.
			\item Split $\{X_{J_b}(t),~ t \in \mathcal{T}\}$ by time dimension into $\omega_{2} \geq 1$ (possibly overlapping) subsets. In this paper, we primarily take $\omega_2 = 1$. However, we have also considered $\omega_{2} = 3$, with $T_1 = \{2000Q1, \ldots, 2007Q4\}$, $T_2 = \{2008Q1, \ldots, 2014Q4\}$ and $T_3 = \{2015Q1, \ldots, 2021Q2\}$, to separate out the common shocks after introduction of the euro, during the financial crisis, the sovereign debt crisis, and COVID-19 outbreak. Such approach could be expanded further by considering, e.g., a rolling-window setting for the choice of time periods $T_j$. Meanwhile $\omega_2=1$ reduces to using the full time sample $\mathcal{T}$; 
			\item For every subset $\mathbf X_{J_b}^{(\ell)} := \{X_{J_b}(t), t \in T_k\}$, $\ell = 1, \ldots, \omega_2$, perform hierarchical clustering with \eqref{eq:dist} as the dissimilarity matrix, taking the relevant dissimilarities for indicators $J_b$, and the number of clusters $\omega_3 \geq 1$;
			\item For every variable $v \in J_b$ and every time-period split, denote which other variables $u \in J_b$ are clustered together with $v$.
		\end{enumerate}
		\item For every pair of variables $i, j \in \mathcal{I}$: 
		\begin{enumerate}
		\item For $i \neq j$, estimate probabilities $p_{ij} = p_{ij}(X_i,X_j)$ as the number of cases the variables are grouped in the same cluster, divided by the number of cases both variables are sampled to $J_b,~ b = 1, \ldots, \omega_1$; 
		\item Set $p_{ii} = 1, \forall i \in \mathcal{I}$;
		\item Generate a distance matrix $\mathcal{D} := (1 - p_{ij})_{i,j = 1}^{n}$. \label{en:step2}
		\end{enumerate}
		\item Probability thresholding:
		\begin{enumerate}
			\item For every $i \in \mathcal{I}$, consider a vector $(p_{i1},\ldots,p_{iN})$, where $N = |\mathcal{I}|$; 
			\item \label{en:step3a} Given a probabilistic drop-out threshold value $\omega_4$, define the cumulative proper clustering likelihood: $L_i := \sum_{p_{ij} > \omega_4} p_{ij};$ 
			\item Order the values obtained by step \eqref{en:step3a}, $L_{(1)} \leq \ldots \leq L_{(N)}$. Given a likelihood threshold parameter $\omega_5 \in [0,1]$, drop those indicators $i\in\mathcal{I}$ from the analysis, which correspond to the smallest $\omega_5$-percentage of the observed $L_{(1)} \leq \ldots \leq L_{(N)}$. Denote the remaining indicator set as $ \mathcal{I}_{\omega_5}$; \label{en:step3c}
			\item Adjust the distance matrix $\mathcal{D}$ from step \eqref{en:step2} to leave only the indicators $i \in \mathcal{I}_{\omega_5}$, obtained by step \eqref{en:step3c}, defining $\mathcal{D}^{\mathcal{I}_{\omega_5}} = (1 - p_{ij})_{i,j \in \mathcal{I}_{\omega_5}}$. \label{en:step3}
		\end{enumerate}
		\item Use the distance matrix $\mathcal{D}^{\mathcal{I}_{\omega_5}}$ from step  \eqref{en:step3} and perform hierarchical clustering on the data $\{ X_i(t), i \in \mathcal{I}_{\omega_5}\}$, with the number of clusters specified as $\omega_6$.  This approach will result in clusters $C_1, \ldots, C_{\omega_6}$.
		In our applications, we set $\omega_6 = \omega_3$. 
		\label{en:step4} 
		\item Using the clusters obtained in step \eqref{en:step4}, estimate the corresponding silhouette scores for every variable $i \in \mathcal{I}_{\omega_5}$, and perform the cluster cleaning algorithm with $\kappa = 0$, defined in Section \ref{sec:sil}. This will result in clusters $\tilde C_1, \ldots, \tilde C_{\omega_6}$ with only positive silhouette scores. 
	\end{enumerate}
	
	The probability thresholding works as a penalty allowing to optimize a certain goodness-of-fit statistic for the estimated clusters, given the chosen set of hyperparameters $(\omega_1, \ldots, \omega_6)$.  When combined, step \eqref{en:step3a} controls the cluster strength under a threshold $\omega_4$, while step \eqref{en:step3c} acts as a penalty on the within-cluster dissimilarities. Finally, step \eqref{en:step3} imposes a variance-based restriction. The proposed approach moves uncertain boundary cases into the drop-out pool of business cycles. In this regard, our proposed soft-clustering approach is similar to soft $k$-means (fuzzy) clustering but differs on how we define the gray zone of the drop-out pool.
	
	In the context of this paper, identifying and excluding the ``outlier'' signals from the remaining less synchronous groups can be an interesting aspect for a deeper investigation. In particular cases, the distinction between specific countries can be increasing over time, as suggested by \cite{wortmann2016one} and \cite{ahlborn2018core}. 
	The proposed clustering approach is reasonably flexible when compared with typical ``hard'' clustering methods. We introduce additional control hyperparameters $(\omega_1, \ldots, \omega_6)$ and a specific dropout function. While $\omega_3, \omega_6$ control the shape of the constructed probabilities, dropout allows cleaning the clusters in the similar sense as Sparse PCA, where the dropped-out series can be inspected separately.
	Besides, as argued previously, the pool of dropped-out series maybe interesting to the investors who seek to diversify the investment portfolios, since these variables uncover the cycles going against the synchronized segments and regions or do not experience visible cyclical behaviour.

	\section{Soft-clustering the European sectoral-regional business cycles}
	\subsection{Overview of extracted business cycles}
	This section summarizes the salient features of the European sectoral-regional business cycles extracted applying the wavelet approach (see Section \ref{sec:wavelet}). We present an analysis in order to provide a deeper understanding of the extraction and contraction phases, average duration and amplitude of any single cycle and give the initial view of the cycles going to a single cluster. 
	
		\begin{figure}[ht!]
		\includegraphics[width = 1\linewidth]{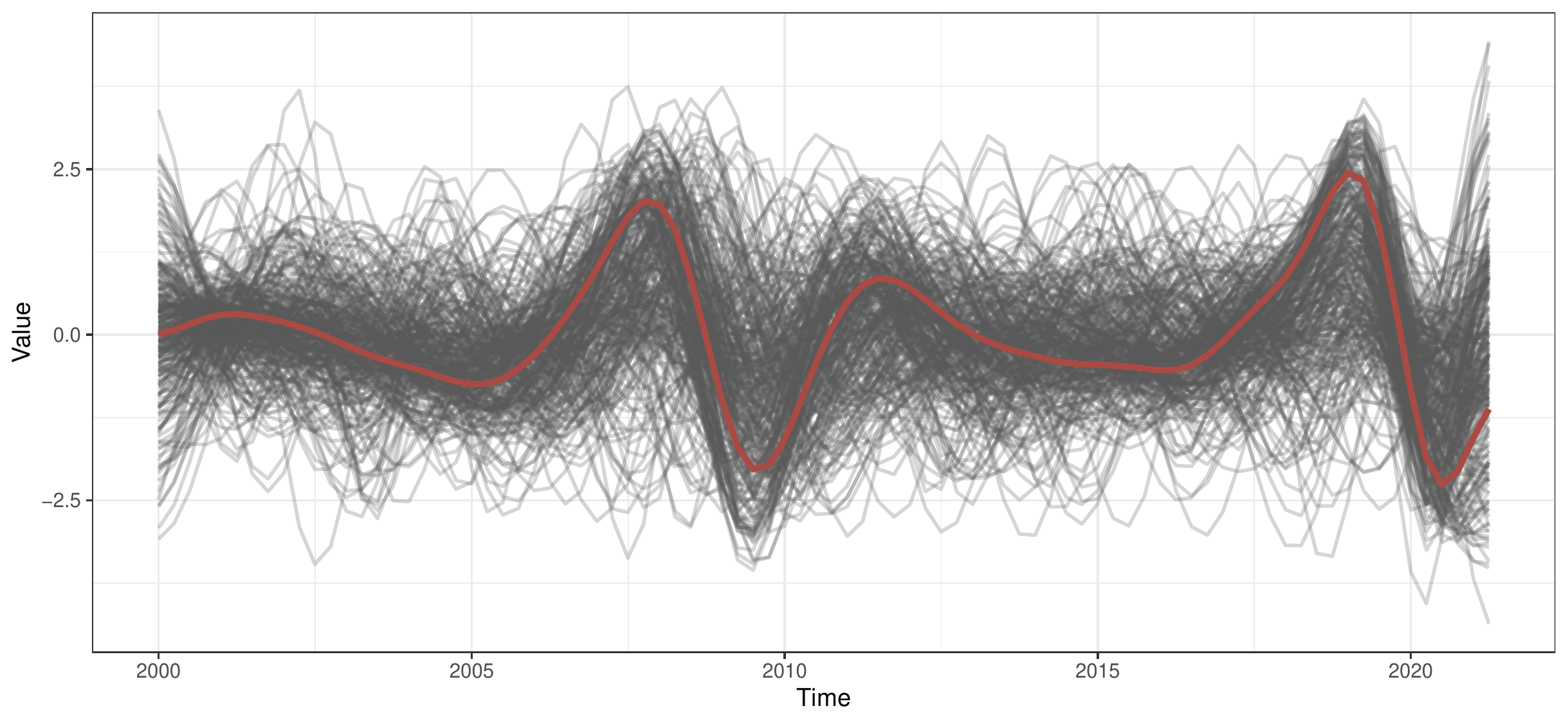}
		\caption{Overview of all of the extracted cycles, with the first principal component in red.} \label{fig:1}
	\end{figure}
	
Figure~\ref{fig:1} demonstrates the single pool of business cycles, together with the first principal component extracted from the scaled data. The principal component analysis produces a low-dimensional representation of the data by finding a sequence of linear combinations of variables, maximizing explained variance. In this case the presented first component explains 32.51\% variance proportion. This suggests the existence of a strongly synchronous cluster within the pool of cycles, implying that the soft-clustering approach may improve the core signal recovery.

\begin{table}[th!]
\centering
\begin{scriptsize}
\caption{Average statistics summarized across sectors and regions.}\label{tab:sum2}
\begin{tabular}{l|rrrrrrrr}

\hline
&Amplitude&Amplitude&Duration&Duration&Cycle&Scale&Synchronicity&Proximity\\
&Expansion&Contraction&Expansion&Contraction&length&&&synchronization\\
\hline
\multicolumn{9}{l}{\textbf{Sectors}}\\ 
\cline{1-9}
A&2.50&2.50&6.93&7.07&14.00&4.98&0.01&-0.10\\ 
B-E&2.42&2.42&8.76&8.21&16.97&3.52&0.45&0.04\\ 
C&2.42&2.40&8.91&8.17&17.08&4.21&0.49&0.05\\ 
F&2.46&2.51&9.44&8.07&17.50&5.67&0.29&0.01\\ 
G-I&2.31&2.42&10.22&7.96&18.18&3.48&0.59&0.08\\ 
J&2.27&2.33&8.52&7.76&16.28&3.04&0.20&-0.04\\ 
K&2.26&2.34&8.09&7.63&15.72&3.17&0.17&-0.06\\ 
L&2.47&2.50&8.92&8.15&17.07&2.00&0.20&-0.04\\ 
M-N&2.35&2.47&9.70&8.01&17.71&3.56&0.51&0.04\\
O-Q&2.26&2.45&9.04&8.33&17.37&1.07&0.07&-0.04\\ 
R-U&1.93&1.86&9.26&8.03&17.29&4.34&0.41&0.05\\ 
\cline{1-9}
\multicolumn{9}{l}{\textbf{Regions}}\\ 
\cline{1-9}
AT&2.13&2.17&8.27&7.35&15.62&2.34&0.51&0.05\\ 
BE&2.44&2.51&8.64&8.75&17.40&2.34&0.28&0.01\\ 
BG&2.33&2.35&7.90&7.78&15.68&3.71&0.24&0.00\\ 
CY&2.46&2.55&9.34&8.35&17.69&3.29&0.25&-0.04\\ 
CZ&2.46&2.39&8.82&7.49&16.31&3.43&0.25&-0.04\\ 
DE&2.33&2.37&9.33&8.17&17.50&3.03&0.41&0.01\\ 
DK&2.23&2.14&8.25&7.29&15.55&2.81&0.24&-0.01\\ 
EE&2.55&2.64&9.48&9.03&18.51&5.69&0.28&0.02\\ 
EL&2.53&2.44&9.54&8.10&17.63&5.28&0.01&-0.07\\ 
ES&2.15&2.24&10.08&7.93&18.00&2.44&0.41&0.03\\ 
FI&2.50&2.45&8.38&8.49&16.86&2.48&0.39&0.02\\ 
FR&2.07&2.21&8.40&7.70&16.10&2.29&0.39&0.03\\ 
HR&2.34&2.33&9.06&7.18&16.25&2.39&0.33&0.01\\ 
HU&2.09&2.19&8.01&6.86&14.87&3.78&0.16&-0.02\\ 
IE&2.18&2.34&9.91&8.87&18.78&6.85&0.30&-0.02\\ 
IT&2.22&2.34&9.37&8.29&17.65&2.21&0.43&0.00\\ 
LT&2.35&2.34&9.13&8.07&17.20&4.40&0.17&0.01\\ 
LU&2.22&2.25&7.08&6.87&13.95&3.88&0.01&-0.06\\ 
LV&2.67&2.63&10.53&8.10&18.63&6.82&0.32&0.02\\
MT&2.07&2.27&7.96&6.83&14.80&2.10&0.22&-0.01\\
NL&2.48&2.49&10.61&9.07&19.68&1.90&0.39&-0.00\\ 
PL&2.08&2.12&7.72&7.59&15.31&3.44&0.16&-0.02\\ 
PT&2.12&2.32&8.06&7.56&15.62&1.99&0.37&0.00\\ 
RO&2.58&2.66&8.46&8.60&17.06&6.35&0.20&-0.03\\ 
SE&2.40&2.36&9.11&8.13&17.24&2.54&0.40&-0.00\\ 
SI&2.49&2.40&8.65&8.64&17.28&2.77&0.40&0.02\\ 
SK&2.44&2.53&7.77&7.24&15.01&5.23&0.22&-0.04\\ 
\cline{1-9}
\multicolumn{9}{l}{\textbf{Total}}\\ 
\cline{1-9}
EU27&2.33&2.38&8.85&7.94&16.80&3.55&0.30&-0.00\\ 
\hline
\end{tabular}
\end{scriptsize}

\end{table}

	For the triangular dissection of the business cycle phases, we use \cite{bbq} quarterly (BBQ) approach as applied in \cite{celovcomunale2021}. The approach locates the local extrema and applies the censoring rules, ensuring that phases alternate between expansion and contraction; phases have a minimum duration of four quarters; and the complete cycle length is at least eight quarters.

	The duration of a phase and the amplitude of changes between alternating peaks and troughs are two sides of the right triangle, the hypotenuse of which shows the average constant growth rate of the phase. The average duration and amplitude allow comparing any pairs of cycles or group averages, examining the symmetry of the stages and computing the average cycle length -- the sum of the arithmetic means of contractions and expansions. The analysis, however, omits the incomplete end-of-sample phases.
	
Table~\ref{tab:sum2} summarizes the sample means of the BBQ approach and includes the average scaling factors, denoting the higher raw amplitudes having groups of business cycles. The table also has the average synchronicity measure \eqref{eq:varphi_2} with the median business cycle taken as the reference point, in line with \cite{mink2012measuring}, and the robust proximity of the synchronization measure \eqref{eq:js1}. 

Although the amplitudes, on average, are symmetric, meaning that a similar size recovery follows the contraction phase, the amplitudes scaling factors vary significantly across sectors and regions. The public sector (O-Q) reacts the least to economic shocks in line with the average semi-elasticities of the cyclically-adjusted EU general government balances being well below unity, while construction (F) and agriculture, forestry and fishing (A) respond the most. On the regional level, the Baltic states, Greece, Romania, and Ireland form the EU periphery that responds the most to the economic shocks, while most core countries have relatively small scaling factors. 

The analysis of durations demonstrates that the data has, on average, from 1 to 3 quarters longer expansion phases than contraction, with more aligned patterns seen on sectoral and diverse observed on regional levels. In some exceptional cases, the regional level shows symmetric phase durations for Slovenia, Romania, Bulgaria and Belgium, implying higher regional diversity and suggesting that further signal refinement may be achieved by soft-clustering on the sectoral-regional level. The average cycle lengths group is around 17 quarters and, on the sectoral level, ranges from 14 quarters for the A sector to 18.2 quarters for trade, hotels, food and transportation (G-I), while, on the regional level, it varies from 14.8 quarters for Malta to 19.7 for the Netherlands.

Finally, we observe high average numbers on sectoral synchronicity levels for industry (B-E), including manufacturing (C), traditional services (G-I), and R\&D-related services (M). These sectors score high on both synchronization measures and suggest that these economic activities potentially form the core of the European synchronous cycle. Regional synchronization diversity hints at the weak evidence for the core-periphery split between the European economies, with the highest scores in Austria, Germany, Spain, Italy, France, and the  Netherlands. The segment with the least synchronous economies includes Greece, Luxembourg, Cyprus, and Visegrád Four. Concerning the core-periphery story, we expect a more discernible split recovered by applying the soft-clustering approach.
	
	\subsection{Choosing soft-clustering parameters}\label{sec:hyperchoice}
	
In this section, we present the soft clustering results considering the full time horizon $\mathcal{T}$, with the number of bootstrapped samples $\omega_1 = 1000$ and $\omega_2 = 1$. Supporting the choice of $\omega_1$, \cite[p. 275]{EfroTibs94} recommend the number of bootstrap samples to be above 500 or 1000 and demonstrate that larger number of samples $\omega_1$ only marginally improves the quality of obtained distributions. Hence, we take the larger of the two numbers. Besides, our experiments with three proposed periods' split ($\omega_2 = 3$), when judged jointly, align with the parsimonious total period view suggesting stability of the synchronous cluster composition. Importantly, the resulting view is not the same as when judging the composition utilizing only one period at once -- this might be crucial for detecting dynamic changes in the composition of the synchronous core cycle or providing event-dependent view. From the standpoint of long-lasting economic policy, a stable composition identification is preferable.

\begin{table}[ht!]
	\centering \scriptsize
	\caption{Asynchronous cluster size.}\label{tab:a3}
	\begin{tabular}{l|cccccccccccc}
		\hline
		\multirow{2}{*}{$\omega_4$}&
		\multicolumn{12}{c}{$1 - \omega_5$} \\ 
		&0.2&0.3&0.4&0.5&0.55&0.6&0.7&0.75&0.8&0.85&0.9&0.95\\ \hline
		0&17&26&54&64&66&69&76&85&96&108&119&123\\
		0.05&17&26&58&64&65&72&77&85&96&108&119&123\\
		0.1&17&26&40&70&72&75&78&85&96&108&119&123\\
		0.15&17&27&39&51&74&76&82&86&96&108&119&123\\
		0.2&39&30&39&74&78&79&85&69&96&108&119&123\\
		0.25&39&33&38&73&76&79&82&77&96&93&119&123\\
		0.3&19&30&37&63&71&70&74&85&96&108&119&123\\
		0.35&17&29&36&55&66&67&72&78&96&104&119&123\\
		0.4&14&28&46&53&57&62&72&80&96&104&103&123\\
		0.45&16&46&46&50&52&55&73&77&91&104&122&123\\
		0.5&32&35&38&42&49&58&73&77&91&104&117&132\\
		0.55&10&10&17&18&28&38&58&71&82&101&122&109\\
		0.6&19&1&8&27&37&51&79&91&102&116&123&109\\
		0.65&17&5&13&32&45&60&87&100&114&122&121&138\\
		0.7&35&5&14&39&53&65&91&102&112&119&129&139\\
		0.75&38&5&20&47&58&71&92&99&100&122&130&131\\
		0.8&37&29&24&52&65&76&92&102&111&121&130&130\\
		0.85&37&5&34&59&65&75&93&103&113&124&126&139\\
		0.9&22&14&41&55&65&75&94&101&107&116&129&126\\
		0.95&1&28&45&53&65&71&92&96&111&120&128&128\\\hline
	\end{tabular}
\end{table}

\begin{table}[ht!]
	\centering \scriptsize
	\caption{Synchronous cluster size.}\label{tab:s2}
	\begin{tabular}{l|cccccccccccc}
		\hline
		\multirow{2}{*}{$\omega_4$}&
		\multicolumn{12}{c}{$1 - \omega_5$} \\  &0.2&0.3&0.4&0.5&0.55&0.6&0.7&0.75&0.8&0.85&0.9&0.95\\ \hline
		0&39&55&44&49&63&72&94&98&98&98&123&144\\
		0.05&39&55&41&49&69&69&81&98&98&98&123&144\\
		0.1&39&55&73&37&53&64&80&98&98&98&123&144\\
		0.15&39&54&73&67&54&66&76&97&98&98&123&144\\
		0.2&18&55&74&34&46&61&71&139&98&98&123&144\\
		0.25&16&53&72&47&45&59&75&122&98&144&123&144\\
		0.3&38&55&73&63&58&73&96&98&98&98&123&144\\
		0.35&40&53&75&73&63&75&98&122&98&122&123&144\\
		0.4&42&54&61&67&75&71&98&121&98&122&147&144\\
		0.45&37&39&45&72&79&86&85&121&122&122&98&144\\
		0.5&26&30&59&85&90&92&85&121&122&122&123&123\\
		0.55&50&74&72&125&130&136&143&144&144&144&127&148\\
		0.6&35&76&111&121&126&127&129&131&135&136&140&147\\
		0.65&36&80&106&116&118&118&121&122&123&127&139&126\\
		0.7&15&83&105&109&110&113&117&120&123&130&118&125\\
		0.75&21&83&99&101&105&107&116&122&128&130&117&144\\
		0.8&21&57&95&96&98&102&116&120&126&113&114&140\\
		0.85&22&84&85&89&98&103&114&119&96&99&134&95\\
		0.9&38&75&78&93&98&103&114&118&125&129&99&143\\
		0.95&59&61&68&95&98&88&114&115&120&101&97&97\\ \hline
	\end{tabular}
\end{table}
	
\begin{figure}[ht!]
		\includegraphics[width = 0.5\linewidth]{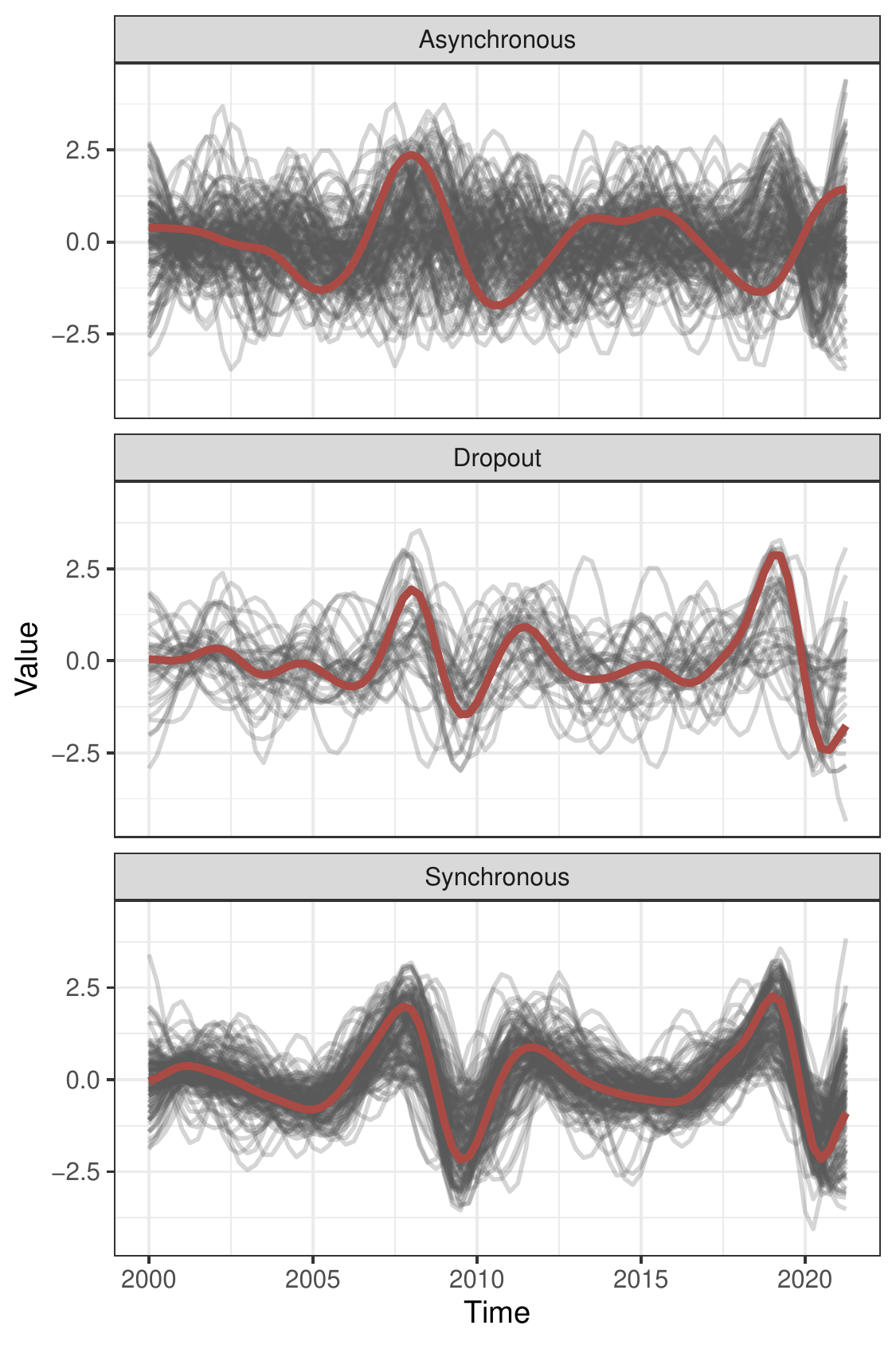}
				\includegraphics[width = 0.5\linewidth]{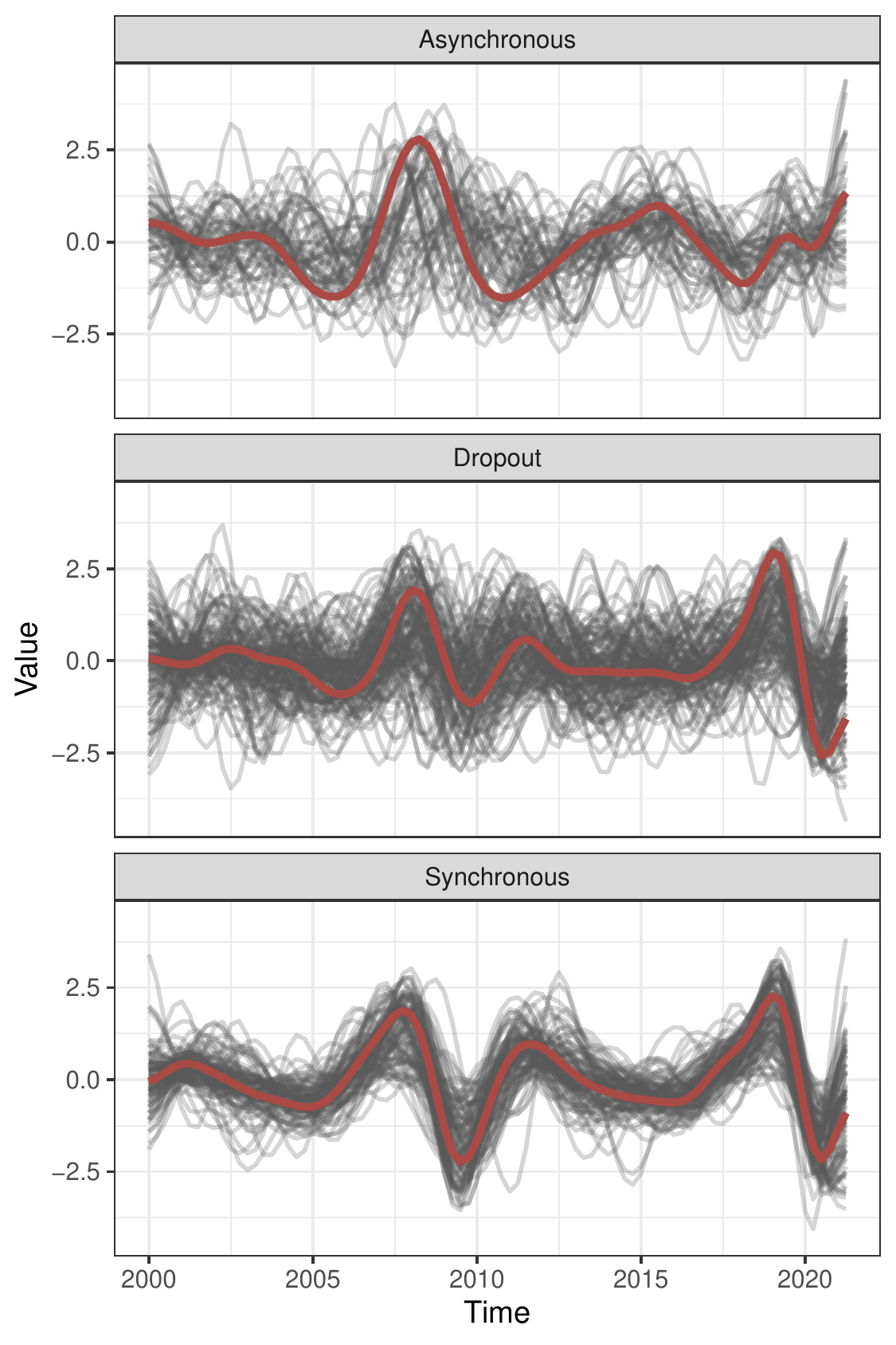}
		\caption{Clustering results: $\omega_4 = 0.8$, with $1 - \omega_5 = 0.95$ (\textbf{left}) and $1 - \omega_5 = 0.55$ (\textbf{right}). Red line denotes the first principal component within the cluster.} \label{fig:2}
	\end{figure}

	In order to find the optimal values for the number of clusters $\omega_3$, the threshold probabilities $\omega_4$ and the dropout rate $\omega_5$, we employ a grid-search approach and examine the resulting cases under different combinations $(\omega_3, \omega_4, \omega_5)$. Appendix~\ref{sec:grid} presents the grid-search results, where we compare the average and minimum 
	silhouette scores pre- and post-cleaning of the clusters, as described in Section \ref{sec:sil}. We seek to select hyperparameters so that the resulting cluster configuration would result in high overall silhouette scores and the weakest resulting cluster wouldn't be ``too weak''. The idea is that the less synchronous cluster can collect all of the noise variables. Such results would infer less synchronization between the variables or that any observed co-movements are complex -- the possible occurrence of phase-shifting or change in frequency dynamics, which fall beyond the scope of business cycle synchronization analysis. Therefore, an alternative solution would be to pool the most asynchronous cycles out of the clusters. Note, there is one caveat about maximizing the silhouette scores. Since, by construction, we drop a chosen fraction of variables based on their observed co-movement strength, the average and minimum scores will increase with increased dropout rates. The trade-off is to eliminate as few variables from the data as possible and find a configuration that retains the highest average silhouette scores. This impact follows from analyzing the resulting cluster sizes in Tables~\ref{tab:s2} and \ref{tab:a3} and the corresponding cluster composition of selected cases in Figure~\ref{fig:2}.
	
	Based on the silhouette scores analysis we note that the most compromising choice for the number of clusters $\omega_3$ is 2. The split into two underlying clusters and a drop-out pool generates the highest silhouette scores continuously over different values of  $(\omega_4, \omega_5)$ (see Tables~\ref{tab:c1}--\ref{tab:c4}). Assumption of such signal composition is consistent with the business cycle synchronization literature (e.g., the core-periphery hypothesis). 
	Besides, due to the specifics of the synchronization measure \eqref{eq:dist} used to form clustering dissimilarities, separating the cycles into two clusters can be seen as a split of the noise from the signal. In other words, we observe that one group consists of highly synchronized business cycles based on the measure \eqref{eq:varphi_1}, and another cluster contains less synchronous series as shown in Figure~\ref{fig:2}. Further analysis of the ``noisy'' group (for $\omega_3 > 2$) reveals smaller synchronizing clusters with different phases, durations and amplitudes than the largest synchronizing group of business cycles. The more detailed splits result in the deterioration of average silhouette scores and advocate for higher dropout rates that does not fit the purpose of our research. Primarily interested in European (euro area) business cycle synchronization, below we elaborate on the results based on the most synchronous cluster from the two, paying less attention to the information from the ``noisy'' group or its further splits. Besides, the clean-up procedure generates the pool of the asynchronous business cycles that may be of further interest for those seeking to recognize potential regions and sectors for optimal investment diversification.  
	
	The other two parameters, $\omega_4$ and $\omega_5$, jointly define the number of discarded business cycles, which form the most asynchronous group of business cycles from the dropout pool. By construction, $\omega_4$ controls the threshold probabilities used in forming the distance matrix of the soft-clustering approach \eqref{en:step3a}, while $\omega_5 \in [0,1]$ denotes the percentile drop-out rate \eqref{en:step3c}. Note, in certain cases we present values for $(1 - \omega_5)$ when highlighting the upper bound percentage of variables remaining in the corresponding analysis. 
	
	Applying the additional adjustments, we discard the noisy series that do not tend to group during the soft-clustering iterations. Setting sufficiently high values for $\omega_4$ reduces the amount of noisy data included in both clusters. From grid-search results provided in Appendix~\ref{sec:grid}, we observe that the compromise average silhouette scores arise when $0.55 \leq \omega_4  \leq 0.95$, and $0.25 \leq \omega_5 \leq 0.65$ with visible nonlinear dependence between the two parameters. Noteworthy, higher number of clusters $\omega_3 > 2$ tends to shift $\omega_5$ closer to $0.8$ drop-out rate, while keeping $\omega_4$ either at values above $0.8$ or below $0.55$. 
	
	Tables~\ref{tab:a3} and \ref{tab:s2} show that increasing $\omega_5$ first affects the less synchronous cluster, leaving the number of business cycles in the main cluster almost unaffected, when $\omega_4$ is between $0.55$ and $0.8$. Therefore, when applying the soft-clustering approach we should pay attention to the trade-off between the size of clusters and the silhouette scores that, at some point, improve because of the asynchronous cluster improvements.
	
	Following the above arguments, we argue that setting $\omega_4$ to higher levels, e.g., $\omega_4 = 0.8$ allows us to adjust the clustering results aiming to extract more consistently synchronized clusters throughout the whole feature space $t \in \mathcal{T}$. In other words, setting $\omega_4 = 0.8$ assumes that the remaining series are grouped at least 80\% of the time during the iterations of the soft-clustering algorithm, indicating the tightness around the mean cycle and suggesting that any dropped variable from the clusters fails to show strong enough bond with the underlying core cycle. Moreover, varying the values of $\omega_4$ allows us to further separate the resulting clusters based on their grouping strength, for instance, establishing weakly co-moving sectors and separate them from strongly co-moving sectors. 
	
\begin{figure}[ht!]
	\includegraphics[width = 1\linewidth]{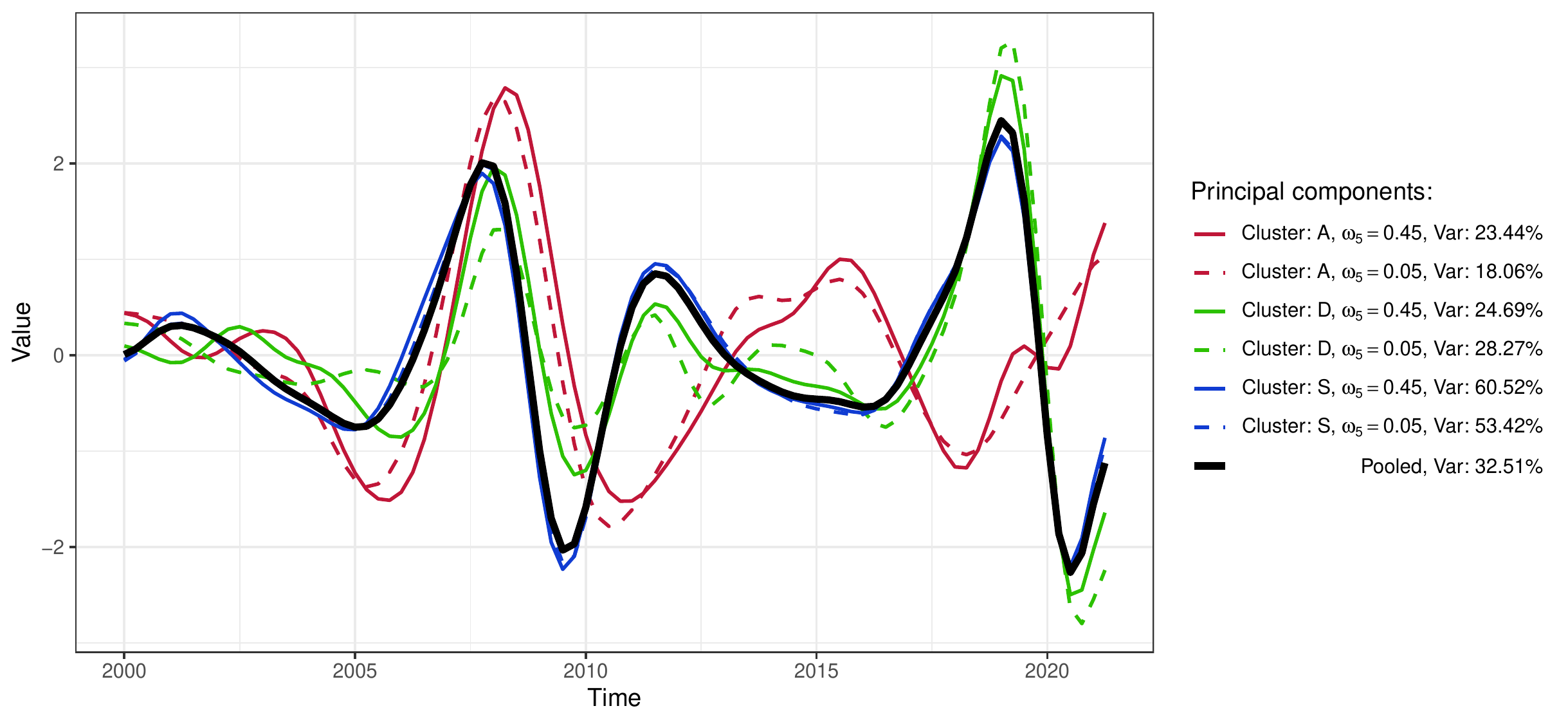}
	\caption{Comparison of the clustering results: first principal components of every formed cluster, $\omega_4 = 0.8$, $\omega_5 \in \{0.05, 0.45\}$, together with the first principal component of all the cycles used in the analysis (in black, denoted as ``Pooled'') and Var showing proportion of variance explained. All principal components are normalized and formed by maximizing the explained variance (varimax rotation).} \label{fig:3}
\end{figure}

The threshold parameter $\omega_5$ controls the number of dropouts from the total data. Note that the dropout is performed based on the $\omega_4$-induced ordering. Thus, the algorithm should begin dropping the series from the ``noisy'' group first and foremost, as seen in both Table~\ref{tab:a3} and Figure~\ref{fig:2}, while only marginally affecting the underlying synchronous cluster. Further analysis of Figures~\ref{fig:2} and \ref{fig:3} reveals that increasing dropout $\omega_5$ from 0.05 to 0.45 improves the synchronization in all three groups. After soft-clustering, the principal components of the synchronous group correspondingly explain the  53.42\% and 60.52\% of variance, and the cluster's composition is less sensitive to the dropout parameter increase. The reshuffle between the asynchronous group and the dropout pool is more evident. The principal components are almost unchanged for the asynchronous cluster, while the composition of the dropout pool is less aligned. Around COVID-19 events, the dropout pool shows the strong synchronization among its components and synchronous cycles. At the same time, the asynchronous cluster moves against the common European business cycle and demonstrates the delayed response to the global financial crisis. Based on the grid-search results, we conclude that setting the $\omega_5$ parameter between $0.45$ and $0.65$ is a reasonable trade-off when applied to the European business cycle problem and, in the empirical part of the paper, set $\omega_5 = 0.45$

	\subsection{The composition of the synchronous European cycle}\label{sec:eu_comp}
	
\begin{figure}[ht!]
	\includegraphics[width=1\linewidth]{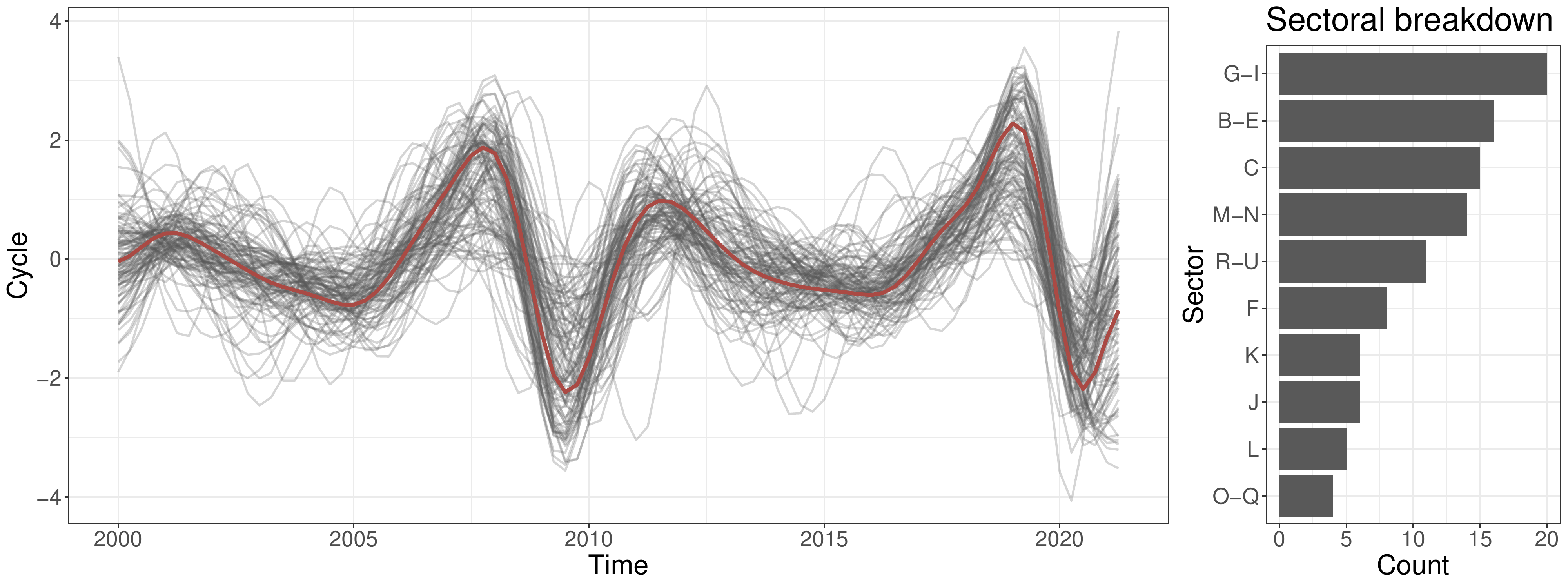}
	\newline
	\includegraphics[width = 0.5\linewidth]{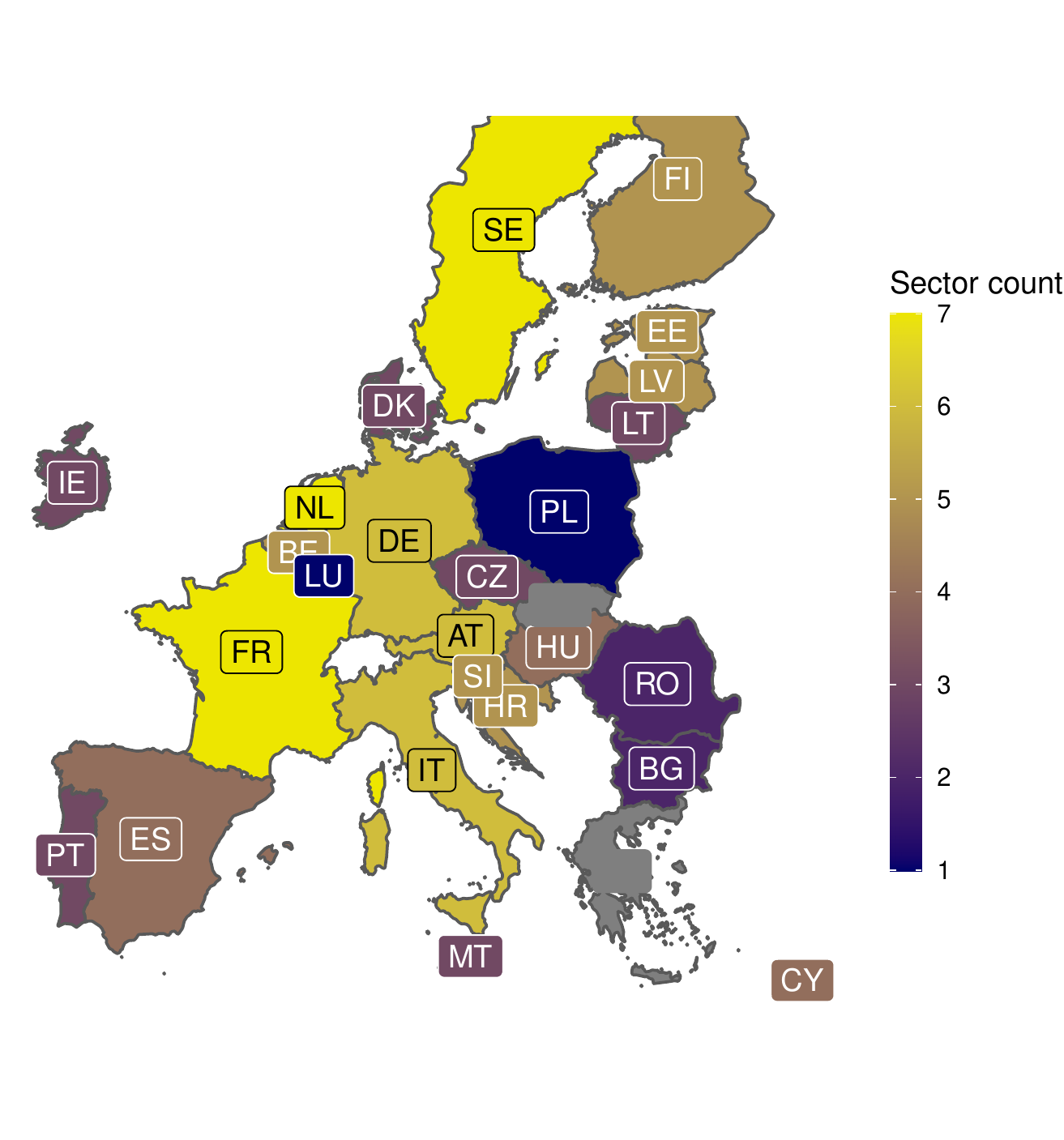}
	\includegraphics[width = 0.5\linewidth]{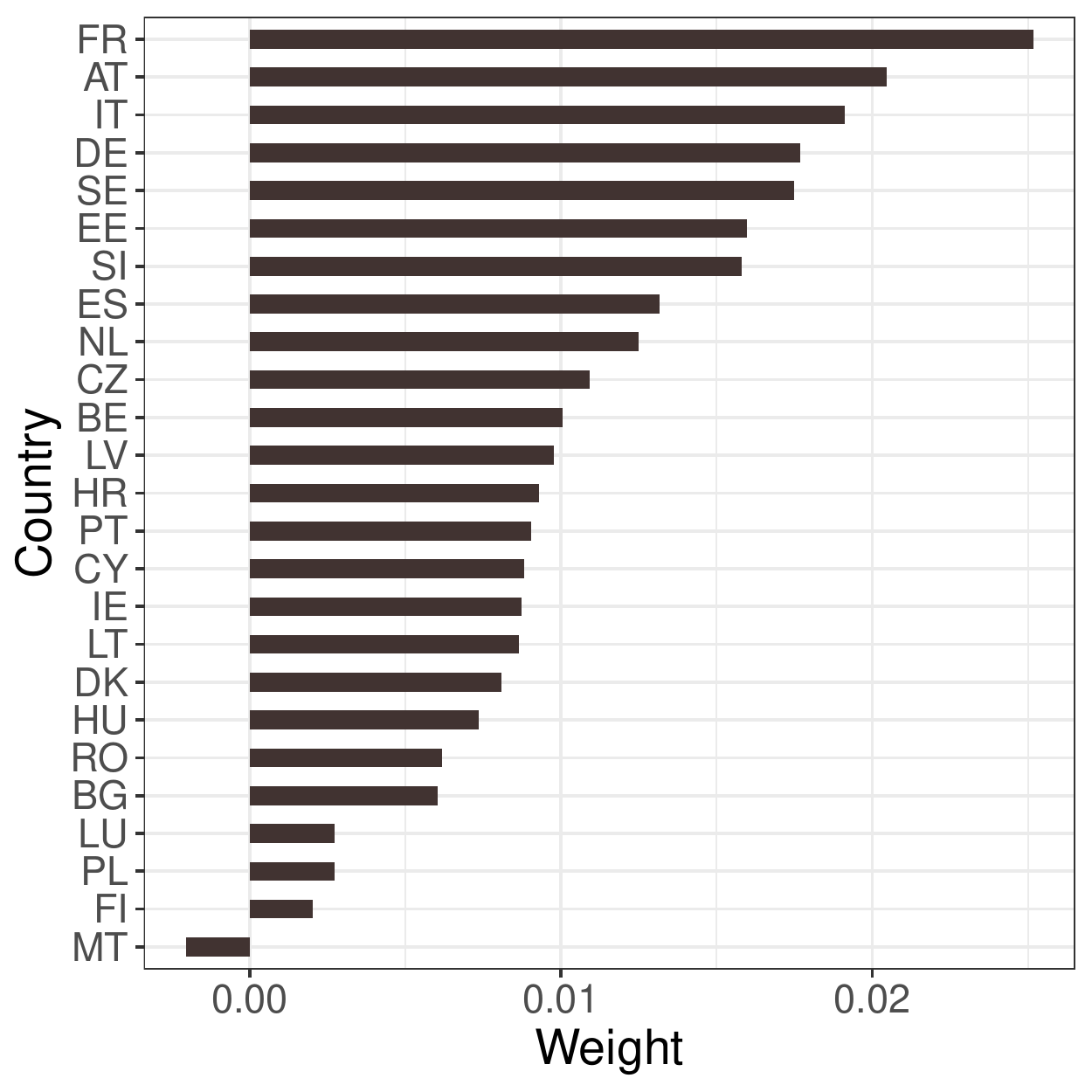}
	\caption{Summary of synchronous cluster ($\omega_4 = 0.8$, $1 - \omega_5 = 0.55$). Overview of the clustered cycles (top left; gray) with the corresponding first principal component (top left; red), sectoral breakdown (top right); map representation of the clustered series, where the color corresponds to the number of different sectors selected from the same country (bottom left); summed proximity of synchronization scores at the country level (bottom right). }\label{fig:4}
\end{figure}

After carefully choosing the soft-clustering parameters in section~\ref{sec:hyperchoice}, we find that parameter setting (1000, 1, 2, 0.8, 0.45, 2) provides a well-balanced view of the synchronous European sectoral-regional set of business cycles. Figure~\ref{fig:4} summarizes the underlying features of the set, while Table~\ref{tab:es} adds information on which sectors comprise the synchronous, asynchronous clusters and the dropout pool.
	
	The upper left graph of Figure~\ref{fig:4} depicts the stacked view of normalized business cycles, where the bold red solid line corresponds to the first principal component computed from the cluster members. The first principal component defines the European business cycle with the highest  60.52\% common variability explained. Judging by the allocation of peaks and troughs, the European business cycle is essentially the same as the signal extracted from the single pool, as demonstrated in Figure~\ref{fig:3}. Hence, the crucial difference is the cleaner view of the main driving sectoral-regional business cycles that constitute the synchronous cluster.

	\begin{table}[ht!]
	\centering
	\caption{Clustering results: $\omega_4 = 0.8$, $1 - \omega_5 = 0.55$. The results consists of two clusters (Asynchronous and Synchronous, denoted by `-' and `+', respectively). The remaining series after the dropout are denoted by blank space in the table and are aggregated under the Dropout cluster.}\label{tab:es}
	\begin{tabular}{l|lllllllllll|rrr}
		\hline
		country & A & B-E & C & F & G-I & J & K & L & M-N & O-Q & R-U & A & S & D \\ 
		\hline
		AT &   & + & + &   & + & + &   &   & + & + &   &   0 &   6 &   5 \\ 
		BE &   & - &   & + &   &   & + & + & + &   & + &   1 &   5 &   5 \\ 
		BG & - & + &   & - & + &   &   & + &   &   &   &   2 &   3 &   6 \\ 
		CY &   &   &   & + & + &   &   &   & + & - &   &   1 &   3 &   7 \\ 
		CZ & - & + & + & - & + &   &   &   &   & - &   &   3 &   3 &   5 \\ 
		DE & - & + & + &   & + & + &   &   & + &   & + &   1 &   6 &   4 \\ 
		DK &   & + & + & - & + & - &   & - &   & - &   &   4 &   3 &   4 \\ 
		EE & - & + & + & + & + & - & - &   & + & - & + &   4 &   6 &   1 \\ 
		EL &   &   &   & - & - & - & - & - &   & - &   &   6 &   0 &   5 \\ 
		ES &   & + & + & - & + &   & + &   & + &   &   &   1 &   5 &   5 \\ 
		FI & - &   &   & + &   &   & + &   & + &   & + &   1 &   4 &   6 \\ 
		FR & - & + & + & + & + & + &   &   & + & + &   &   1 &   7 &   3 \\ 
		HR & - &   & + & - & + &   & + &   &   & + & + &   2 &   5 &   4 \\ 
		HU & - & + & + &   &   &   & - & + & + &   &   &   2 &   4 &   5 \\ 
		IE & - &   &   &   & + & - &   &   &   & + & + &   2 &   3 &   6 \\ 
		IT &   & + & + & + & + &   & + &   &   & - & + &   1 &   6 &   4 \\ 
		LT & - & + & + &   & + & - &   & - &   &   &   &   3 &   3 &   5 \\ 
		LU &   & - & - & - &   &   & - &   &   & - &   &   5 &   0 &   6 \\ 
		LV & - & + & + &   & + & + & - &   &   & - & + &   3 &   5 &   3 \\ 
		MT &   & - &   & - &   & + &   & - & + & - & - &   5 &   2 &   4 \\ 
		NL & - & + & + &   & + & + &   & - & + &   & + &   2 &   6 &   3 \\ 
		PL & - &   &   &   & + &   &   & - &   & - &   &   3 &   1 &   7 \\ 
		PT & - & + & + &   & + &   &   &   &   &   &   &   1 &   3 &   7 \\ 
		RO &   &   &   & - & + & - &   &   &   &   & + &   2 &   2 &   7 \\ 
		SE &   & + & + & + & + &   & + &   & + &   & + &   0 &   7 &   4 \\ 
		SI &   & + & + & - & + &   & - & - &   &   & + &   3 &   4 &   4 \\ 
		SK &   &   &   &   &   &   & - &   &   & - &   &   2 &   0 &   9 \\ \hline \hline
		A & 14 & 3 & 1 & 10 & 1 & 6 & 7 & 7 & 0 & 11 & 1 &  &  &  \\ 
		S & 0 & 16 & 16 & 7 & 20 & 6 & 6 & 3 & 12 & 4 & 12 &  &  &  \\ 
		D & 13 & 8 & 10 & 10 & 6 & 15 & 14 & 17 & 15 & 12 & 14 &  &  &  \\ 
		\hline
	\end{tabular}

\end{table}

The allocation of peaks and troughs of the European business cycle adheres to the most prominent global economic shocks observed in the recent two decades. After 2000, the dot-com bubble burst recovers relatively fast with visibly less synchronous levels among the components and other groups (see Figure~\ref{fig:3}). On the back of the non-prudentially booming credit market, the underlying European cycle slowly moves to the overheating expansion, which speeds up after the enlargement of the EU in 2004. Consequently, the overheated EU market gets a heavy punch from the global financial crisis of 2007-2009, with a double deep recovery due to the subsequent European sovereign debt crisis. Finally, the pressures on the EU labor market resulted in a too hot position for the EU economies before the COVID-19 outbreak. However, since the latest wing is close to the end of the sample, we judge its severity with caution.

    The upper right graph of Figure~\ref{fig:4} shows the sectoral composition of the synchronous cluster. Supplemented by the numbers in Table~\ref{tab:a3}, the diagram demonstrates that the core signal encompasses the business cycle originating from traditional services, including trade, hotels, food and transportation (G-I). In addition, the core contains the industry in the narrow (C) and broad (B-E) sense and R\&D related services (M-N). Tourism-related arts, recreation, entertainment and other services (R-U) demonstrate mixed results. However, since the statistical institutions often use this segment as a recycle bin in balancing the GDP definition by production, expenditure and income approaches, the R-U sector requires careful treatment. The large proportion (14 out of 27) of R-U going to the dropout pool supports this view. 
	
	The lower left part of Figure~\ref{fig:4} shows the regional contributions to the European business cycle, where the color corresponds to the number (count) of different economic activities of the same country. We argue that the higher number of sectors in the synchronous cluster demonstrates a stronger core signal, hence, increases the contribution of such economies in the synchronous group. Supplemented by the numbers in Table~\ref{tab:a3}, the observed results suggest that the core signal originates from the main contributors to the EU budget: France, Sweden, Netherlands, Germany, Austria, Italy, Belgium and Spain, followed closely by small and open Estonia's economy. In this context, Estonia's business cycle is a barometer of the European business cycle synchronization with the core countries. On the other end is the broad spectrum of periphery countries, with no contribution from Greece and Luxembourg that are the central players in the asynchronous cluster, while Slovakia adding the most to the dropout pool.

	On the other hand, the analysis of the asynchronous group and the dropout pool in Table~\ref{tab:es} equips investors with a set of economic activities and regions least synchronized with the underlying European economic cycle. Both groups contain all business cycles for agriculture, forestry and fishing (A), and the significant part of public services (O-Q) that also experience the lowest fluctuations (see, Table~\ref{tab:sum2}). The groups include financial and insurance (K) and information and telecommunication (J) services with some additional impact stemming from construction (F). On the regional profile, the less synchronous group are driven by the EU outliers: Luxembourg with the core activities and Greece with the asynchronous cluster-specific activities. The dropout pool is more aligned with the periphery composition, including heavily hit by sovereign debt crisis Cyprus, Portugal, Ireland, Finland, Spain, and Malta, and nearly all data for Slovakia with, however, less evident reasons for the inclusion. In sum, the asynchronous pool is a mixture of safe-haven activities and regions with their opposite -- regions and sectors that are not recommended for investments during recessions.

The results in Figure~\ref{fig:4} support our previous statement that France, Italy, Austria, Germany, The Netherlands, Sweden, and Spain form the regional core of the cluster, and Estonia serves as a barometer. The remaining European countries show weaker synchronization both in counts and in proximity measures, with Malta likely being the least synchronous with the rest of the countries and some countries unable to compute due to zero count of business cycles in the synchronous group.

\subsection{Discussion of core-periphery hypothesis}
	The core-periphery results presented in the previous section closely align with the observations from the related literature. We find that the approach taken in this paper adds a level of granularity, that helps defining the underlying synchronization structures. In this section we elaborate further on the resulting evidence, confirming certain known results. Furthermore, we distinguish and discuss the arising disparities from the perspective of the sectoral composition of the core.
	
First, we find evidence that France, together with Spain and Germany, forms the core cycle. These results are similar to \cite{belo2001some}. France is seen as the leader of the common euro area business cycles, e.g., by \cite{aguiar2011oil} and \cite{kapounek2019historical}, among others. However, regarding Germany, the results in the literature are more mixed, with \cite{ahlborn2018core}, \cite{kapounek2019historical}, \cite{bunyan2020fiscal}, \cite{celovcomunale2021} observing the decoupling of the German business cycles from the EU area. Our results suggest that this is not the case when we consider the main open economic activities, including broad industry (B-E), traditional services (G-I), IT and telecommunication (J), and R\&D (M-N). Hence, we conclude that soft-clustering helps strengthen Germany's core signal by removing other ``noisy'' economic activities.
	
	Second, we find that Greece is not included in the cluster, suggesting evidence of weak synchronization with the growth cycles of the core. Such observation results are consistent with the literature (see, e.g., \cite{bunyan2020fiscal}, \cite{campos2016core}). Similarly to \cite{campos2016core}, we find that Ireland, Portugal, and Greece, which are argued to form a smaller periphery by the authors, are also weakly present in the core cluster. The only inconsistency is that the authors suggest that Spain should also belong to the smaller periphery with the aforementioned regions, which is contradicted by our soft-clustering view, while missing Cyprus and, especially, Malta. At the same time, Luxembourg economy is another, wealthier end outlier with business cycles not in the core and not widely discussed in the literature.
	
	As expected by \cite{landesmann2003ceecs}, \cite{MONFORT2013689}, \cite{bandres2017regional}, a clear differentiation between regions exists, where old EU member states may diverge from the economies joined after 2004 and experiencing fast catching-up/overheating episodes, for instance, for the Baltic States, and very small representation in the core of Greece, Bulgaria, Romania, and Visegrád Four.

\section{Concluding remarks}

To sum up, the sectoral-regional look at the development of European economies in the recent decades opens up prospects for analysing a broad range of problems. From a macroeconomic point of view, it is crucial to understand which economic activities and regions drive the European business cycle and which lag in responding to the shocks or even move in the opposite direction. In this regard, our paper confirms the existing core-periphery findings, keeping the main contributors to the EU budget in the synchronous core, of which all but Sweden are the members of the euro area. The core regions are strongly supported by open-to-trade industrial production, trade, food, accommodation and transportation services, and R\&D economic activities -- a hidden aspect when just looking at the regional level. 

Having a stable synchronous core is a necessary condition for successful OCA development. However, we find that not all euro area members enter the synchronous cluster. The latest adopters of the euro, small and open economies of the Baltic states, are improving in their catch-up with the synchronous core in recent decades. Furthermore, Central European entrants, Greece, Cyprus, Malta, Ireland, and Luxembourg contribute to the asynchronous cluster and the dropout data pool. These findings support the previously observed core-periphery regional division by adding agriculture, public sector, IT, finance and insurance, real estate, and other services to the asynchronous group of economic activities and the dropout pool.

From an econometric point of view, we argue the importance of tidying the synchronous core signal. Definitely, both the wavelets used for the trend-cycle decomposition and the new soft-clustering approach proposed in the paper have opportunities for further development. 

For a more thorough analysis and as a robustness check, the interested reader may consider alternative trend-cycle decomposition methods, for instance, \cite{RePEc:fip:fedcwp:9906} for the Christiano-Fitzgerald filter, \cite{phillips2019boosting} for a boosted HP filter and \cite{LEE2020} for sparse HP filter, among many others. \cite{celovcomunale2021} provides a comprehensive Monte-Carlo simulation experiment comparing various trend-cycle decomposition methods, including the wavelet approach. In the earlier versions of the paper, alongside the MODWT approach, we also worked with continuous wavelets, Singular Spectrum Analysis (SSA) and Complete Ensemble Empirical Mode Decomposition with Adaptive Noise (CEEMDAN, \cite{5947265}). These methods are decent alternatives to the wavelet approach due to their flexibility and a good fit for the frequency domain analysis. However, the focus of the current version of the paper is on improving the soft-clustering approach, leaving the aforementioned alternative approaches for future work. Furthermore, when considering further possible avenues for wavelet application, we find promising results by considering half-peak/trough reflection in order to better gauge the end-of-sample problems.

The soft-clustering approach is already pretty flexible to adjust for tidying the cluster of interest in a data-rich environment. The method employs bootstrapped sampling, probabilistic thresholding and silhouette scores for cleaning the cluster and removing any unstable, in the chosen dissimilarity metrics sense, observations. As the results depend on the dissimilarity metric, we conclude that its choice must stem from the problem analyzed. In this sense, business cycle synchronization perfectly matches the use of synchronicity measures. 

In the paper, we attack the stability aspect from several angles. Besides the traditional choice of the number of clusters, bootstrapped sampling facilitates reshuffling the data and highlights the elements in the stable core, together with the boundary cases. These boundary data points tend to migrate from sample to sample, reducing the probability of sitting together with all other data in the same cluster. In this regard, probabilistic thresholding is the crucial clean-up procedure that improves the visibility of the synchronous core and highlights sectors and regions that must come first when focusing on the European economic policy. For future work, it would be interesting to delve deeper into the estimation of optimal probabilistic thresholding parameter values. Although the soft-clustering method belongs to the realm of unsupervised learning approaches, we deem that for a particular problem (convergence, synchronization), it is possible to find the problem-specific objective function. Solving the problem under particular (e.g., sparsity-synchronicity) trade-off restrictions, we could find the optimal combination of the parameters. Finally, we can scrutinize the dynamic composition of the synchronous core by applying, for instance, the rolling-window approach.

\section*{Acknowledgements}
This research has received funding from the European Social Fund (project No. 09.3.3-LMT-K-712-01-123) under a grant agreement with the Research Council of Lithuania (LMTLT).

The authors would like to thank Svatopluk Kapounek, Jesus Crespo Cuaresma and all participants of “Euro4Europe” workshops for comments and suggestions. 

% \bibliography{reference}

\newpage

\begin{appendices}

\section{Comparison of synchronous business cycle composition}

The analysis below demonstrates the trade-off between highlighting the core sectoral-regional data points and the size of the dropout rates. We find that the small drop-out rates tend to keep many boundary cases, while too high values remove the business cycles from the core open-to-trade sectors of the economy too much. The core regions, however, are the least affected. Hence, we could suggest higher drop-out rates for highlighting the main contributors but moderate adopting the findings when formulating the economic policy recommendations.

\begin{table}[H]
	\centering
	\footnotesize
	\caption{Composition of synchronous business cycles clusters under differing drop-out rates. Here the probability threshold is 80\% and drop-out rates are 5\%, 45\%, 65\%. The `*, **, ***' denote the inclusion at different drop-out rates, with `***' meaning inclusion at all three cases, and `*' inclusion only under 5\% drop-out.}
\begin{tabular}{c|ccccccccccc}
	\hline
	\multirow{2}{*}{Country} & \multicolumn{11}{c}{Sector}\\
	 & A & B-E & C & F & G-I & J & K & L & M-N & O-Q & R-U \\ 
	\hline
	AT & * & *** & *** & * & *** & *** &   & * & *** & *** & * \\ 
	BE &   &   &   & *** &   &   & *** & *** & *** &   & *** \\ 
	BG &   & *** & * &   & *** &   &   & *** &   &   &   \\ 
	CY &   &   &   & *** & *** &   &   & * & *** &   & * \\ 
	CZ &   & *** & *** &   & *** &   &   &   & * &   &   \\ 
	DE &   & *** & *** &   & *** & ** &   &   & *** &   & *** \\ 
	DK &   & *** & *** &   & *** &   &   &   &   &   &   \\ 
	EE &   & *** & *** & *** & *** &   &   &   & *** &   & ** \\ 
	EL &   & * &   &   &   &   &   &   &   &   &   \\ 
	ES &   & *** & *** &   & *** & * & * & * & *** &   &   \\ 
	FI &   & * & * & *** & * &   & ** & ** & *** &   & *** \\ 
	FR &   & *** & *** & *** & *** & *** &   &   & *** & *** & * \\ 
	HR &   & ** & ** &   & *** & * & *** &   & * & *** & *** \\ 
	HU &   & *** & *** &   & * &   &   & *** & *** &   & * \\ 
	IE &   &   &   & * & *** &   &   &   &   & *** & *** \\ 
	IT &   & *** & *** & *** & *** &   & *** & * & ** &   & *** \\ 
	LT &   & *** & *** &   & ** &   &   &   & * & * & * \\ 
	LU &   &   &   &   & * &   &   &   & *** &   &   \\ 
	LV &   & *** & *** & * & *** & *** &   &   & * &   & *** \\ 
	MT &   &   & * &   &   & *** & * &   & *** &   &   \\ 
	NL &   & *** & *** & ** & *** & *** &   &   & *** &   & *** \\ 
	PL &   &   &   &   & *** &   &   &   & * &   &   \\ 
	PT &   & *** & *** &   & *** &   & * & * &   &   &   \\ 
	RO &   & * & * &   & *** &   &   &   & * &   & *** \\ 
	SE &   & *** & *** & *** & *** &   & *** &   & *** &   & *** \\ 
	SI &   & *** & *** &   & *** &   &   &   & ** &   & *** \\ 
	SK &   &   &   & * & * & ** &   &   &   &   &   \\ 
	\hline
\end{tabular}
\end{table}

\section{Alternative set of parameters}
We argued in the paper that higher drop-out rates have a smaller impact on the synchronous core and demonstrated this by analyzing the optimal, in our view, set of hyperparameters. To complete the discussion, in this Appendix, we show an alternative set of parameters with a much smaller dropout rate $\omega_5 \in \{0.05, 0.65\}$. We notice that all the main conclusions of the paper remain valid.

\begin{figure}[H]
	\includegraphics[width=1\linewidth]{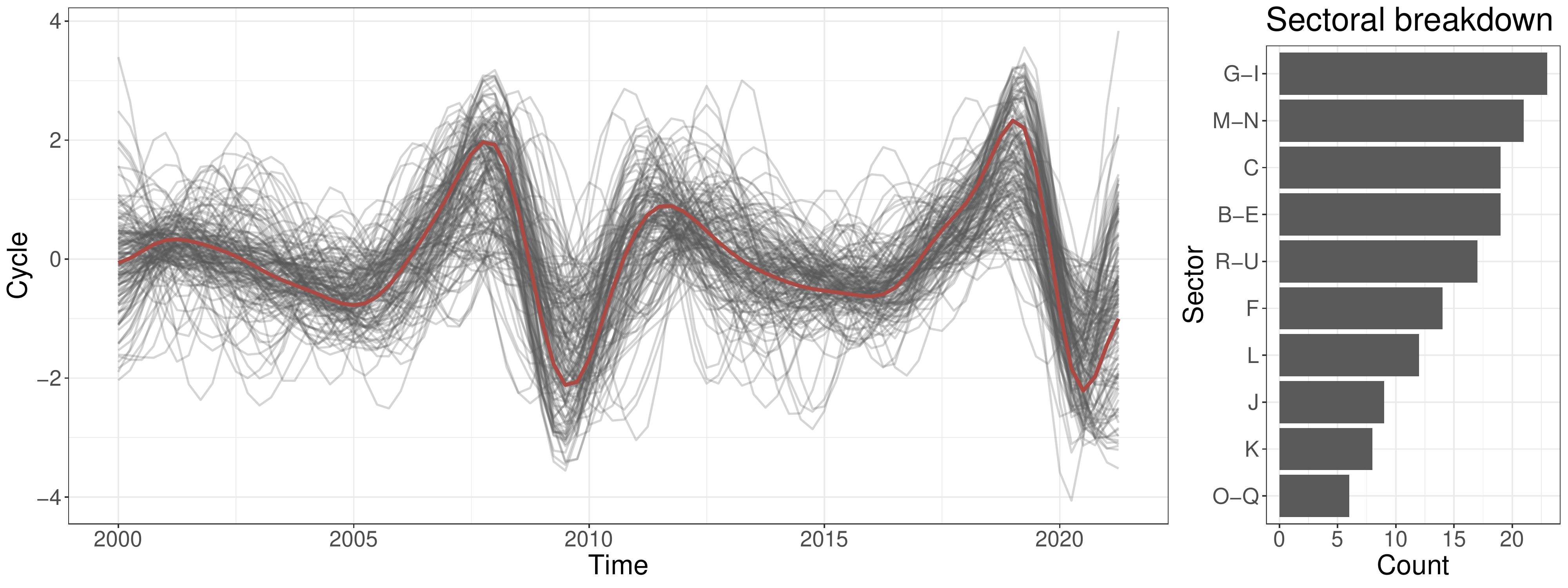}
	\newline
	\includegraphics[width = 0.5\linewidth]{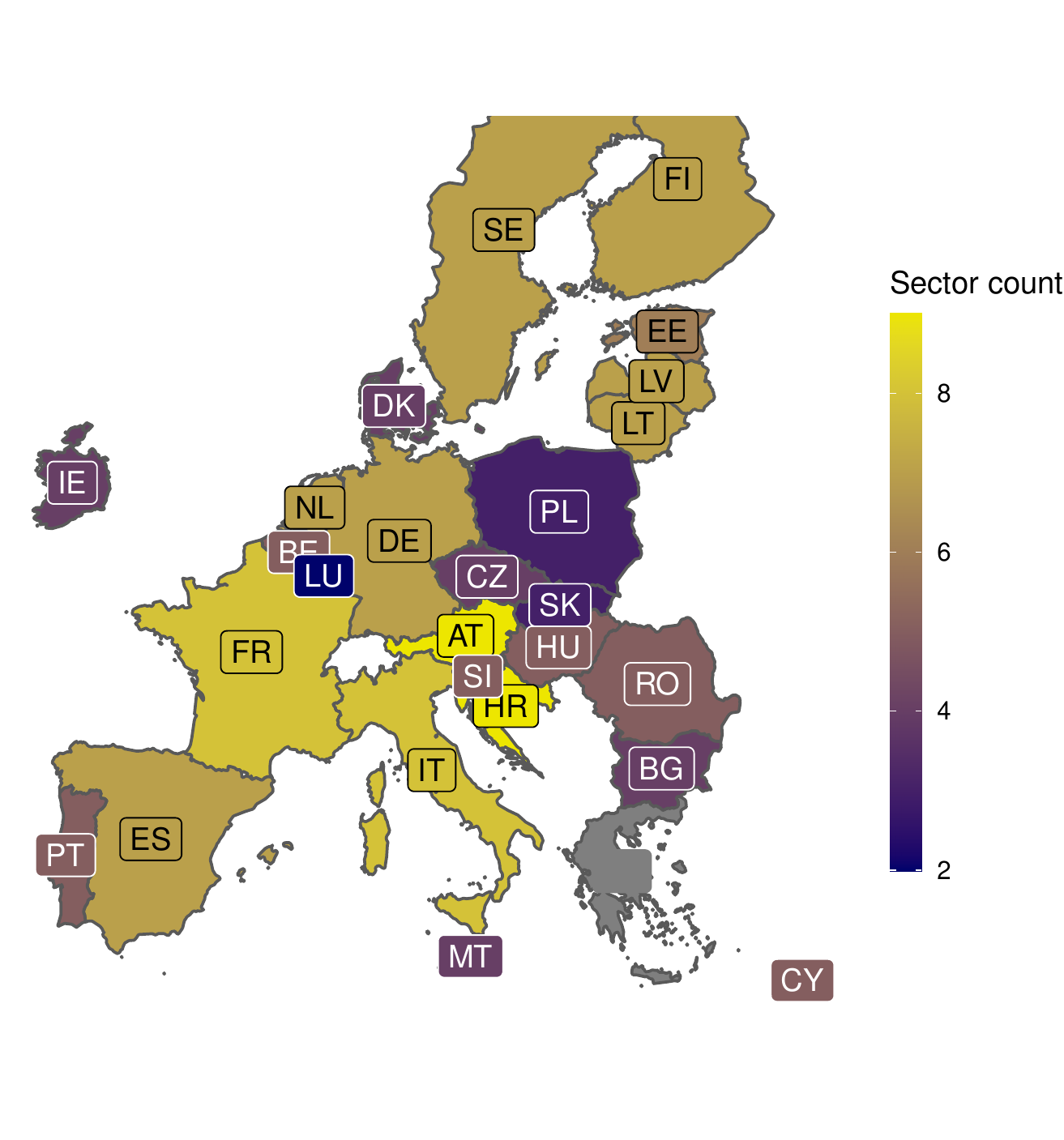}
	\includegraphics[width = 0.5\linewidth]{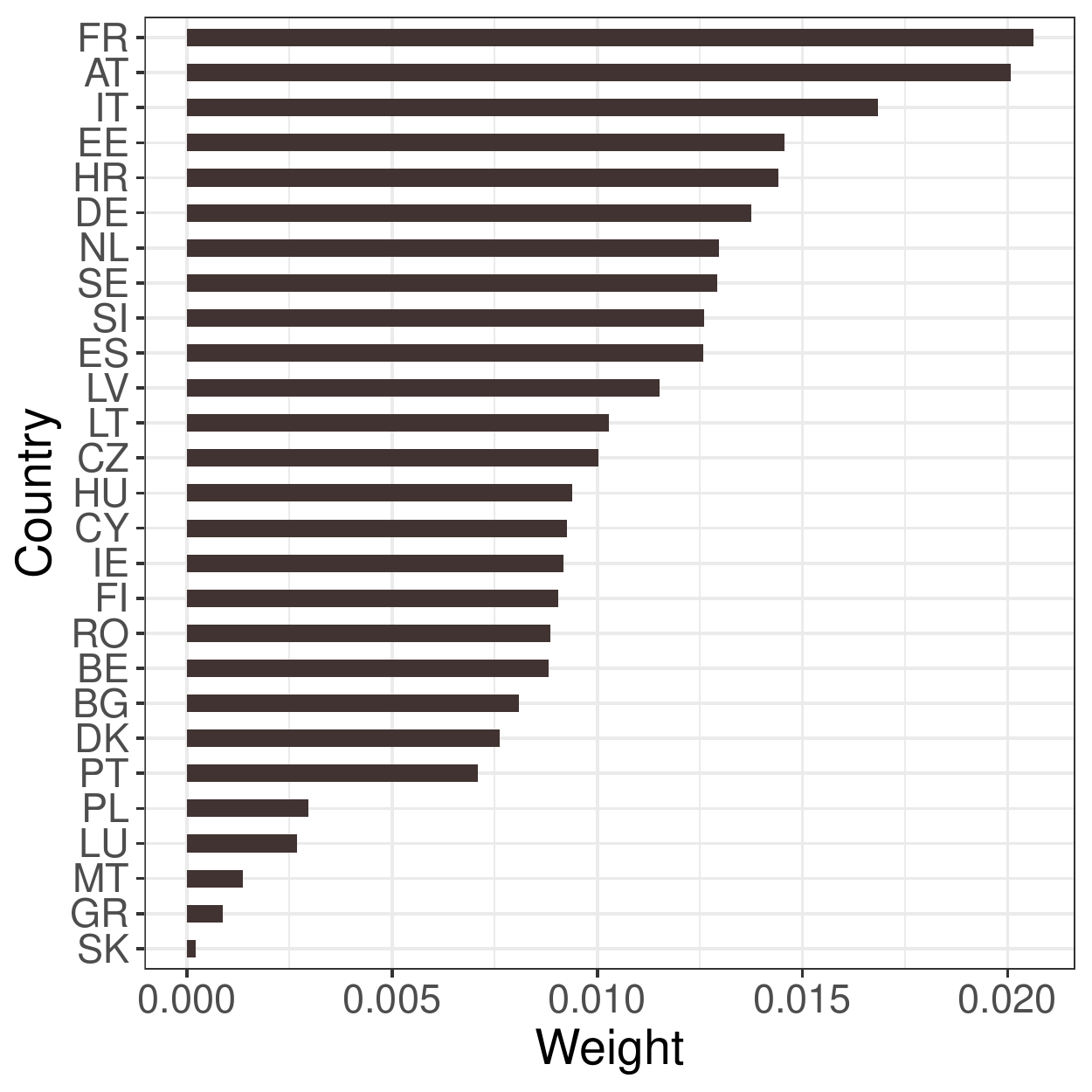}
	\caption{Summary of Cluster 2 ($\omega_4 = 0.8$, $1 - \omega_5 = 0.95$). Overview of the clustered cycles (top left; gray) with the corresponding first principal component (top left; red), sectoral breakdown (top right); map representation of the clustered series, where the color corresponds to the number of different sectors selected from the same country (bottom left); summed proximity of synchronization scores at the country level (bottom right). }
\end{figure}

\begin{table}[H]
	\centering%\scriptsize
	\caption{Clustering results: $\omega_4 = 0.8$, $1-\omega_5 = 0.95$. The results consists of two clusters (Asynchronous and Synchronous, denoted by `-` and `+`, respectively). The remaining series after the drop-out are denoted by blank space in the table and are aggregated under the Dropout cluster.}
	\begin{tabular}{l|lllllllllll|rrr}
		\hline 
		country & A & B-E & C & F & G-I & J & K & L & M-N & O-Q & R-U & A & S & D \\ 
		\hline
		AT &   & + & + &   & + & + & - & + & + & + & + &   1 &   8 &   2 \\ 
		BE & - & - & - & + & - & - & + & + & + & - & + &   6 &   5 &   0 \\ 
		BG & - & + & + & - & + & - & - & + & - & - & - &   7 &   4 &   0 \\ 
		CY &   & - & - & + & + & + & - & + & + & - & + &   4 &   6 &   1 \\ 
		CZ & - & + & + & - & + & - & - & - & + & - & - &   7 &   4 &   0 \\ 
		DE & - & + & + & - & + & + &   &   & + & - & + &   3 &   6 &   2 \\ 
		DK & - & + & + & - & + & - &   & - &   & - & - &   6 &   3 &   2 \\ 
		EE & - & + & + & + & + & - & - & - & + & - & + &   5 &   6 &   0 \\ 
		EL & - & + & - & - & - & - & - & - & - & - & - &  10 &   1 &   0 \\ 
		ES & - & + & + & - & + & + & + & + & + & - & - &   4 &   7 &   0 \\ 
		FI & - & + & + & + & + & - & + & + & + & + & + &   2 &   9 &   0 \\ 
		FR & - & + & + & + & + & + & - & - & + & + & + &   3 &   8 &   0 \\ 
		HR & - & + & + & - & + & + & + & + & + & + & + &   2 &   9 &   0 \\ 
		HU & - & + & + & - & + & - & - & + & + &   &   &   4 &   5 &   2 \\ 
		IE & - & - & - & + & + & - &   & - &   & + & + &   5 &   4 &   2 \\ 
		IT & - & + & + & + & + & - & + & + & + & - & + &   3 &   8 &   0 \\ 
		LT & - & + & + & + & + & - & - & - & + & + & + &   4 &   7 &   0 \\ 
		LU & - & - & - & - & + &   & - & - & + & - &   &   7 &   2 &   2 \\ 
		LV & - & + & + & + & + & + & - & - & + & - & + &   4 &   7 &   0 \\ 
		MT &   & - &   & - & - & + & + & - & + & - & - &   6 &   3 &   2 \\ 
		NL & - & + & + & + & + & + & - & - & + & - & + &   4 &   7 &   0 \\ 
		PL & - &   &   &   & + & - & - & - & + & - &   &   5 &   2 &   4 \\ 
		PT & - & + & + & - & + &   & + & + &   & - & - &   4 &   5 &   2 \\ 
		RO & - &   & + & - & + & - & - & - & + & - & + &   6 &   4 &   1 \\ 
		SE & - & + & + & + & + & - & + &   & + & - & + &   3 &   7 &   1 \\ 
		SI &   & + & + & - & + & - & - & - & + &   & + &   4 &   5 &   2 \\ 
		SK & - & - & - & + &   & + & - & - & - & - & - &   8 &   2 &   1 \\ \hline \hline
		A & 23 & 6 & 6 & 13 & 3 & 15 & 16 & 15 & 3 & 19 & 8 &  &  &  \\ 
		S & 0 & 19 & 19 & 12 & 23 & 10 & 8 & 10 & 21 & 6 & 16 &  &  &  \\ 
		D & 4 & 2 & 2 & 2 & 1 & 2 & 3 & 2 & 3 & 2 & 3 &  &  &  \\ 
		\hline
	\end{tabular}

\end{table}
\begin{figure}[H]
	\includegraphics[width=1\linewidth]{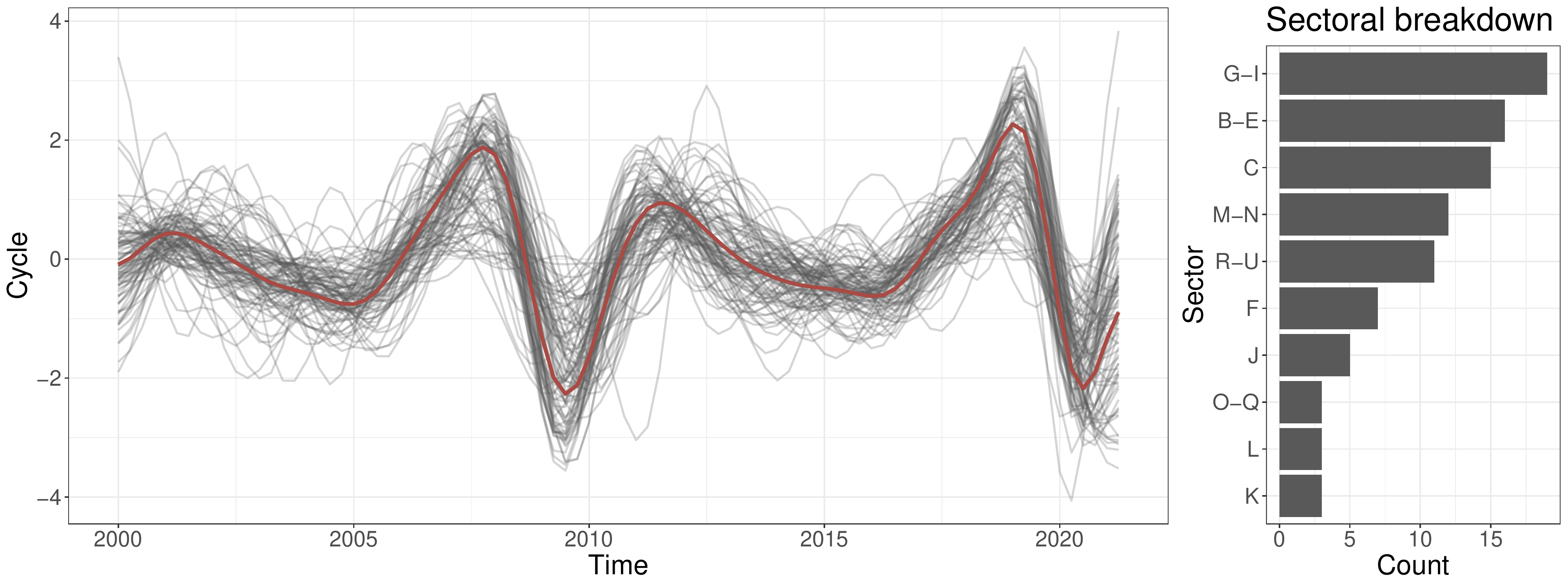}
	\newline
	\includegraphics[width = 0.5\linewidth]{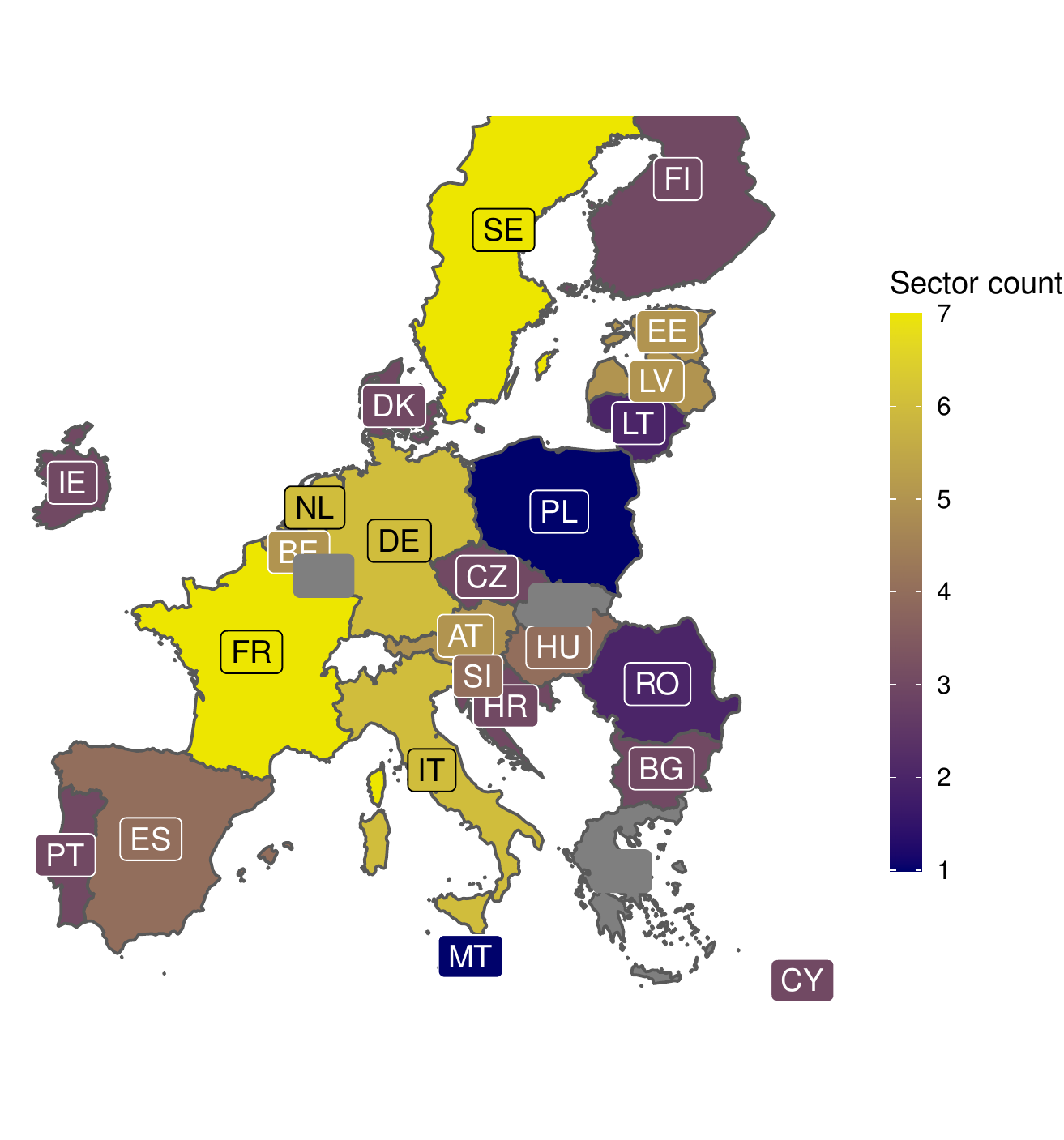}
	\includegraphics[width = 0.5\linewidth]{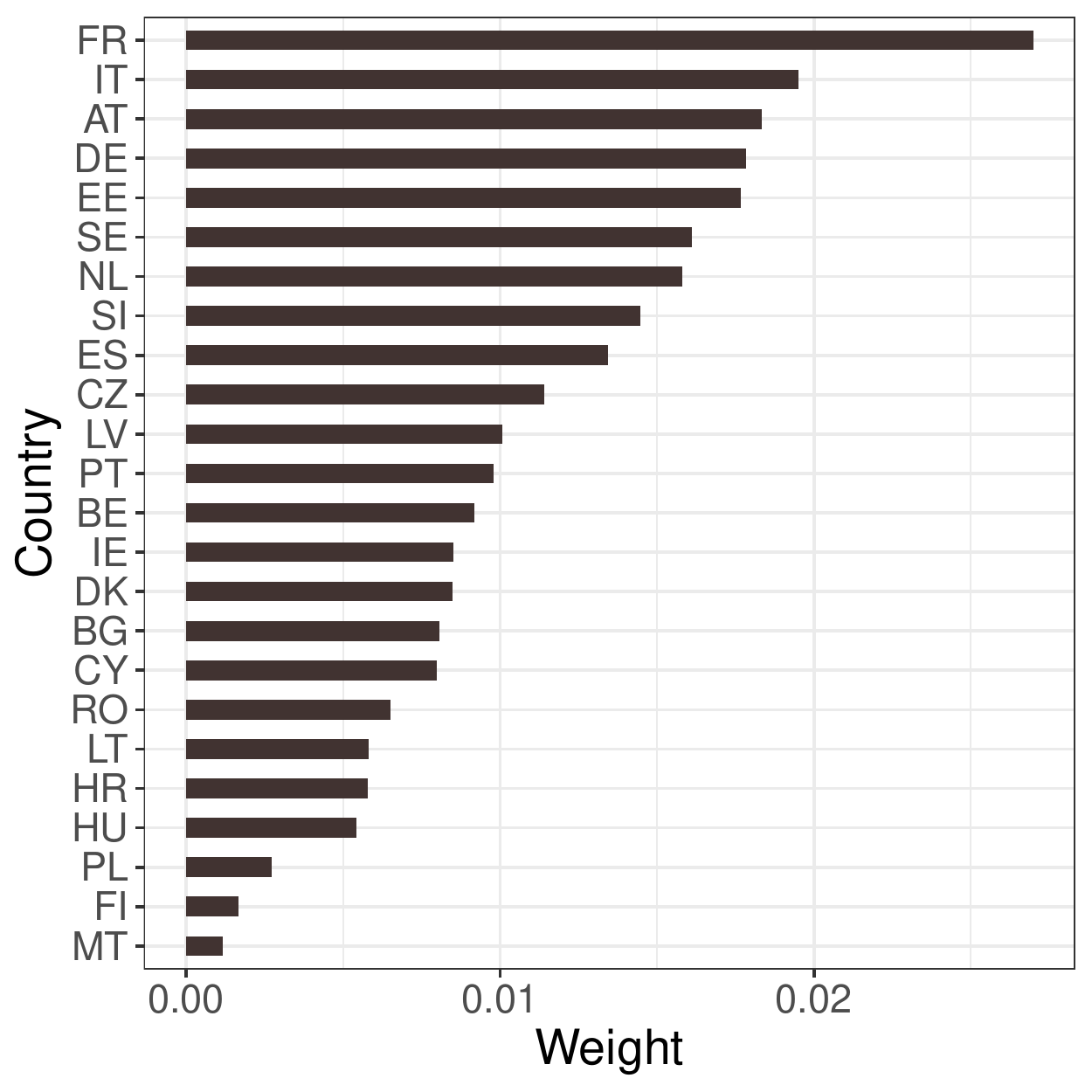}
	\caption{Summary of Cluster 2 ($\omega_4 = 0.8$, $1 - \omega_5 = 0.35$). Overview of the clustered cycles (top left; gray) with the corresponding first principal component (top left; red), sectoral breakdown (top right); map representation of the clustered series, where the color corresponds to the number of different sectors selected from the same country (bottom left); summed proximity of synchronization scores at the country level (bottom right). }
\end{figure}

\begin{table}[H]
	\centering %\scriptsize
	\caption{Clustering results: $\omega_4 = 0.8$, $1 - \omega_5 = 0.35$. The results consists of two clusters (Asynchronous and Synchronous, denoted by `-` and `+`, respectively). The remaining series after the drop-out are denoted by blank space in the table and are aggregated under the Dropout cluster.}
	\begin{tabular}{l|lllllllllll|rrr}
		\hline
		country & A & B-E & C & F & G-I & J & K & L & M-N & O-Q & R-U & A & S & D \\ 
		\hline
		AT &   & + & + &   & + & + &   &   & + & + &   &   0 &   6 &   5 \\ 
		BE &   &   &   & + &   &   & + & + & + &   & + &   0 &   5 &   6 \\ 
		BG &   & + &   &   & + &   &   & + &   &   &   &   0 &   3 &   8 \\ 
		CY &   &   &   & + & + &   &   &   & + &   &   &   0 &   3 &   8 \\ 
		CZ &   & + & + & - & + &   &   &   &   &   &   &   1 &   3 &   7 \\ 
		DE & - & + & + &   & + &   &   &   & + &   & + &   1 &   5 &   5 \\ 
		DK &   & + & + &   & + &   &   &   &   & - &   &   1 &   3 &   7 \\ 
		EE &   & + & + & + & + &   &   &   & + &   & + &   0 &   6 &   5 \\ 
		EL &   &   &   &   &   &   & - &   &   &   &   &   1 &   0 &  10 \\ 
		ES &   & + & + &   & + &   &   &   & + &   &   &   0 &   4 &   7 \\ 
		FI &   &   &   & + &   &   &   &   & + &   & + &   0 &   3 &   8 \\ 
		FR &   & + & + & + & + & + &   &   & + & + &   &   0 &   7 &   4 \\ 
		HR & - &   &   &   & + &   &   &   &   & + & + &   1 &   3 &   7 \\ 
		HU &   & + & + &   &   &   &   & + & + &   &   &   0 &   4 &   7 \\ 
		IE & - &   &   &   & + &   &   &   &   & + & + &   1 &   3 &   7 \\ 
		IT &   & + & + & + & + &   & + &   &   &   & + &   0 &   6 &   5 \\ 
		LT &   & + & + &   &   &   &   &   &   &   &   &   0 &   2 &   9 \\ 
		LU &   & - &   &   &   &   &   &   & + &   &   &   1 &   1 &   9 \\ 
		LV &   & + & + &   & + & + &   &   &   &   & + &   0 &   5 &   6 \\ 
		MT &   &   &   &   &   &   &   &   & + & - &   &   1 &   1 &   9 \\ 
		NL &   & + & + &   & + & + &   &   & + &   & + &   0 &   6 &   5 \\ 
		PL &   &   &   &   & + &   &   &   &   &   &   &   0 &   1 &  10 \\ 
		PT &   & + & + &   & + &   &   &   &   &   &   &   0 &   3 &   8 \\ 
		RO &   &   &   &   & + &   &   &   &   &   & + &   0 &   2 &   9 \\ 
		SE &   & + & + & + & + &   & + &   & + &   & + &   0 &   7 &   4 \\ 
		SI &   & + & + &   & + &   &   &   &   &   & + &   0 &   4 &   7 \\ 
		SK &   &   &   &   &   &   &   &   &   &   &   &   0 &   0 &  11 \\ \hline \hline
		A & 3 & 1 & 0 & 1 & 0 & 0 & 1 & 0 & 0 & 2 & 0 &  &  &  \\ 
		S & 0 & 16 & 15 & 7 & 19 & 4 & 3 & 3 & 13 & 4 & 12 &  &  &  \\ 
		D & 24 & 10 & 12 & 19 & 8 & 23 & 23 & 24 & 14 & 21 & 15 &  &  &  \\ 
		\hline
	\end{tabular}
	
\end{table}

\newpage
\section{Grid-search analysis}
\label{sec:grid}
The appendix summarizes the grid-search results looking for the hyperparameters of the soft-clustering approach $(\omega_3,\omega_4,\omega_5)$, when $\omega_1 = 1000$ bootstrapped samples, $\omega_2 = 1$, and $\omega_6 = \omega_3$. The tables show the average and minimum silhouette scores before (initial) and after silhouette clean-up (final) from Section~\ref{sec:sil}. 

{\scriptsize
\begin{longtable}{c|c|ccccccccccc}\caption{Composition of silhouette scores during the grid-search. Here $\omega_1 = 1000$, $\omega_2 = 1$ and $\omega_6 = \omega_3$. The presented scores correspond to the average score between all clusters, after the silhouette clean-up. }\label{tab:c1} \\  \hline 
		\multirow{2}{*}{$\omega_3$}&\multirow{2}{*}{$\omega_4$}&\multicolumn{11}{c}{$1 - \omega_5$} \\
			&&0.2&0.25&0.35&0.45&0.55&0.65&0.75&0.8&0.85&0.9&0.95\\
		\hline
		\endfirsthead
		
		\caption{\textit{(Continued)} Composition of silhouette scores during the grid-search. Here $\omega_1 = 1000$, $\omega_2 = 1$ and $\omega_6 = \omega_3$. The presented scores correspond to the average score between all clusters, after the silhouette clean-up. } \\  \hline 
		\multirow{2}{*}{$\omega_3$}&\multirow{2}{*}{$\omega_4$}&\multicolumn{11}{c}{$1 - \omega_5$} \\
		&&0.2&0.25&0.35&0.45&0.55&0.65&0.75&0.8&0.85&0.9&0.95\\
		\hline
		\endhead
		2&0&0.33&0.36&0.47&0.58&0.58&0.63&0.59&0.56&0.6&0.59&0.59\\
		2&0.05&0.38&0.35&0.42&0.52&0.56&0.66&0.59&0.64&0.57&0.64&0.63\\
		2&0.1&0.33&0.35&0.39&0.41&0.57&0.65&0.62&0.63&0.57&0.58&0.61\\
		2&0.15&0.35&0.32&0.4&0.45&0.51&0.61&0.63&0.62&0.63&0.57&0.61\\
		2&0.2&0.36&0.33&0.29&0.47&0.48&0.61&0.62&0.63&0.57&0.62&0.6\\
		2&0.25&0.35&0.34&0.28&0.4&0.51&0.57&0.61&0.64&0.55&0.6&0.59\\
		2&0.3&0.39&0.27&0.29&0.44&0.46&0.57&0.59&0.63&0.63&0.61&0.59\\
		2&0.35&0.31&0.29&0.26&0.43&0.49&0.6&0.63&0.63&0.57&0.59&0.63\\
		2&0.4&0.25&0.26&0.48&0.52&0.61&0.66&0.65&0.62&0.62&0.59&0.59\\
		2&0.45&0.29&0.38&0.47&0.58&0.63&0.59&0.63&0.62&0.59&0.61&0.61\\
		2&0.5&0.45&0.61&0.68&0.68&0.65&0.62&0.68&0.66&0.62&0.61&0.61\\
		2&0.55&0.67&0.71&0.71&0.7&0.7&0.62&0.64&0.65&0.66&0.65&0.62\\
		2&0.6&0.61&0.54&0.72&0.77&0.74&0.73&0.72&0.72&0.68&0.67&0.63\\
		2&0.65&0.56&0.62&0.38&0.8&0.8&0.79&0.75&0.72&0.7&0.67&0.63\\
		2&0.7&0.6&0.57&0.88&0.85&0.85&0.81&0.76&0.72&0.68&0.66&0.68\\
		2&0.75&0.59&0.45&0.87&0.87&0.86&0.82&0.75&0.71&0.69&0.69&0.63\\
		2&0.8&0.66&0.61&0.89&0.9&0.86&0.81&0.74&0.7&0.69&0.66&0.65\\
		2&0.85&0.68&0.55&0.93&0.91&0.84&0.78&0.72&0.7&0.66&0.66&0.61\\
		2&0.9&0.66&0.54&0.94&0.87&0.82&0.77&0.72&0.69&0.69&0.64&0.65\\
		2&0.95&0.97&0.97&0.94&0.84&0.76&0.74&0.67&0.66&0.68&0.65&0.67\\ \hline
		3&0&0.67&0.57&0.58&0.52&0.49&0.47&0.43&0.42&0.43&0.41&0.41\\
		3&0.05&0.54&0.53&0.56&0.46&0.51&0.44&0.45&0.46&0.44&0.42&0.42\\
		3&0.1&0.5&0.51&0.49&0.47&0.48&0.41&0.43&0.45&0.43&0.42&0.44\\
		3&0.15&0.59&0.43&0.46&0.45&0.47&0.42&0.46&0.45&0.46&0.42&0.43\\
		3&0.2&0.47&0.55&0.52&0.44&0.45&0.43&0.47&0.42&0.45&0.44&0.43\\
		3&0.25&0.63&0.61&0.56&0.47&0.45&0.43&0.42&0.47&0.45&0.49&0.43\\
		3&0.3&0.66&0.62&0.59&0.49&0.53&0.4&0.45&0.43&0.4&0.5&0.48\\
		3&0.35&0.76&0.73&0.59&0.52&0.45&0.5&0.47&0.43&0.46&0.44&0.44\\
		3&0.4&0.72&0.64&0.55&0.49&0.52&0.51&0.49&0.44&0.45&0.44&0.43\\
		3&0.45&0.74&0.75&0.56&0.49&0.54&0.52&0.48&0.48&0.45&0.43&0.52\\
		3&0.5&0.72&0.64&0.57&0.58&0.61&0.54&0.47&0.47&0.45&0.48&0.46\\
		3&0.55&0.68&0.67&0.56&0.59&0.6&0.53&0.5&0.49&0.52&0.46&0.42\\
		3&0.6&0.75&0.6&0.58&0.6&0.6&0.51&0.48&0.46&0.46&0.45&0.44\\
		3&0.65&0.72&0.59&0.65&0.64&0.59&0.5&0.54&0.47&0.46&0.5&0.44\\
		3&0.7&0.57&0.62&0.67&0.64&0.55&0.48&0.5&0.5&0.48&0.49&0.48\\
		3&0.75&0.48&0.52&0.7&0.63&0.51&0.51&0.5&0.49&0.52&0.44&0.46\\
		3&0.8&0.67&0.67&0.69&0.54&0.57&0.6&0.49&0.51&0.48&0.48&0.46\\
		3&0.85&0.61&0.76&0.67&0.54&0.58&0.61&0.6&0.54&0.5&0.46&0.5\\
		3&0.9&0.8&0.74&0.65&0.59&0.52&0.63&0.58&0.54&0.49&0.49&0.52\\
		3&0.95&0.85&0.8&0.58&0.55&0.58&0.55&0.57&0.61&0.49&0.52&0.6\\ \hline
		4&0&0.66&0.72&0.49&0.47&0.47&0.44&0.43&0.43&0.42&0.41&0.41\\
		4&0.05&0.73&0.66&0.62&0.51&0.46&0.48&0.45&0.45&0.45&0.42&0.41\\
		4&0.1&0.62&0.66&0.46&0.52&0.48&0.45&0.44&0.42&0.43&0.45&0.45\\
		4&0.15&0.69&0.74&0.62&0.56&0.45&0.4&0.43&0.43&0.43&0.43&0.41\\
		4&0.2&0.55&0.72&0.48&0.59&0.48&0.46&0.46&0.44&0.43&0.42&0.42\\
		4&0.25&0.68&0.6&0.64&0.55&0.52&0.53&0.53&0.45&0.46&0.42&0.42\\
		4&0.3&0.61&0.51&0.49&0.57&0.52&0.56&0.52&0.45&0.49&0.42&0.41\\
		4&0.35&0.64&0.57&0.5&0.56&0.53&0.52&0.48&0.48&0.44&0.45&0.43\\
		4&0.4&0.72&0.54&0.54&0.52&0.5&0.55&0.53&0.47&0.47&0.42&0.43\\
		4&0.45&0.57&0.61&0.52&0.57&0.55&0.57&0.48&0.5&0.47&0.43&0.44\\
		4&0.5&0.57&0.65&0.56&0.6&0.58&0.56&0.49&0.45&0.46&0.46&0.43\\
		4&0.55&0.6&0.6&0.55&0.6&0.62&0.55&0.47&0.48&0.47&0.44&0.43\\
		4&0.6&0.49&0.55&0.6&0.65&0.59&0.54&0.49&0.46&0.41&0.42&0.44\\
		4&0.65&0.59&0.59&0.61&0.68&0.59&0.54&0.51&0.46&0.46&0.46&0.44\\
		4&0.7&0.62&0.6&0.62&0.64&0.56&0.54&0.51&0.47&0.47&0.44&0.43\\
		4&0.75&0.62&0.6&0.71&0.64&0.56&0.52&0.47&0.49&0.46&0.46&0.46\\
		4&0.8&0.63&0.66&0.7&0.61&0.57&0.51&0.49&0.45&0.45&0.48&0.44\\
		4&0.85&0.79&0.81&0.64&0.59&0.53&0.5&0.48&0.49&0.44&0.46&0.44\\
		4&0.9&0.86&0.76&0.63&0.56&0.52&0.49&0.45&0.43&0.43&0.44&0.45\\
		4&0.95&0.79&0.71&0.63&0.54&0.52&0.48&0.48&0.48&0.48&0.48&0.49\\ \hline \hline
\end{longtable}
}

{\scriptsize
	\begin{longtable}{c|c|ccccccccccc}\caption{Composition of silhouette scores during the grid-search. Here $\omega_1 = 1000$, $\omega_2 = 1$ and $\omega_6 = \omega_3$. The presented scores correspond to the minimum score between all clusters, after the silhouette clean-up. }\label{tab:c2} \\  \hline 
		\multirow{2}{*}{$\omega_3$}&\multirow{2}{*}{$\omega_4$}&\multicolumn{11}{c}{$1 - \omega_5$} \\
		&&0.2&0.25&0.35&0.45&0.55&0.65&0.75&0.8&0.85&0.9&0.95\\
		\hline
		\endfirsthead
		
		\caption{\textit{(Continued)} Composition of silhouette scores during the grid-search. Here $\omega_1 = 1000$, $\omega_2 = 1$ and $\omega_6 = \omega_3$. The presented scores correspond to the minimum score between all clusters, after the silhouette clean-up. } \\  \hline 
		\multirow{2}{*}{$\omega_3$}&\multirow{2}{*}{$\omega_4$}&\multicolumn{11}{c}{$1 - \omega_5$} \\
		&&0.2&0.25&0.35&0.45&0.55&0.65&0.75&0.8&0.85&0.9&0.95\\
		\hline
		\endhead
	2&0&0.32&0.23&0.25&0.26&0.35&0.36&0.39&0.46&0.46&0.55&0.55\\
	2&0.05&0.36&0.32&0.34&0.3&0.31&0.34&0.4&0.43&0.48&0.47&0.51\\
	2&0.1&0.26&0.23&0.27&0.32&0.32&0.32&0.38&0.41&0.5&0.54&0.55\\
	2&0.15&0.24&0.2&0.34&0.25&0.34&0.35&0.42&0.4&0.43&0.51&0.58\\
	2&0.2&0.22&0.23&0.2&0.3&0.34&0.37&0.39&0.41&0.45&0.46&0.57\\
	2&0.25&0.34&0.22&0.2&0.33&0.35&0.35&0.36&0.43&0.44&0.53&0.54\\
	2&0.3&0.31&0.2&0.2&0.3&0.32&0.33&0.41&0.42&0.44&0.58&0.54\\
	2&0.35&0.22&0.19&0.24&0.26&0.28&0.3&0.36&0.43&0.47&0.54&0.5\\
	2&0.4&0.18&0.19&0.25&0.23&0.24&0.33&0.39&0.43&0.44&0.54&0.5\\
	2&0.45&0.23&0.23&0.27&0.25&0.22&0.28&0.42&0.6&0.5&0.56&0.59\\
	2&0.5&0.18&0.23&0.22&0.2&0.23&0.39&0.46&0.46&0.59&0.58&0.58\\
	2&0.55&0.22&0.23&0.3&0.3&0.7&0.56&0.6&0.64&0.62&0.63&0.6\\
	2&0.6&0.52&0.37&0.47&0.76&0.72&0.68&0.7&0.68&0.64&0.64&0.61\\
	2&0.65&0.36&0.46&0.24&0.8&0.78&0.77&0.71&0.68&0.64&0.64&0.6\\
	2&0.7&0.56&0.42&0&0.84&0.85&0.79&0.71&0.68&0.65&0.63&0.56\\
	2&0.75&0.55&0.41&0&0.87&0.85&0.78&0.71&0.66&0.63&0.61&0.59\\
	2&0.8&0.5&0.49&0.89&0.89&0.85&0.79&0.7&0.66&0.66&0.64&0.64\\
	2&0.85&0.55&0.43&0.92&0.91&0.8&0.74&0.66&0.65&0.62&0.6&0.56\\
	2&0.9&0.5&0.48&0.94&0.86&0.78&0.71&0.63&0.65&0.6&0.56&0.63\\
	2&0.95&0.97&0.96&0.93&0.77&0.69&0.65&0.66&0.64&0.66&0.59&0.65\\ \hline
	3&0&0.56&0.51&0.38&0.31&0.25&0.28&0.26&0.26&0.31&0.31&0.28\\
	3&0.05&0.47&0.49&0.47&0.35&0.3&0.23&0.28&0.25&0.29&0.3&0.31\\
	3&0.1&0.43&0.48&0.35&0.2&0.25&0.17&0.25&0.28&0.28&0.29&0.28\\
	3&0.15&0.22&0.29&0.22&0.24&0.2&0.2&0.25&0.23&0.29&0.34&0.32\\
	3&0.2&0.32&0.38&0.34&0.2&0.18&0.21&0.23&0.3&0.26&0.27&0.3\\
	3&0.25&0.33&0.28&0.29&0.22&0.2&0.19&0.26&0.31&0.31&0.2&0.34\\
	3&0.3&0.2&0.35&0.34&0.19&0.29&0.3&0.32&0.35&0.38&0.25&0.24\\
	3&0.35&0.15&0.49&0.46&0.29&0.41&0.33&0.42&0.32&0.34&0.43&0.22\\
	3&0.4&0.14&0.17&0.39&0.47&0.45&0.42&0.35&0.35&0.4&0.38&0.41\\
	3&0.45&0.14&0.61&0.53&0.41&0.45&0.42&0.43&0.41&0.36&0.42&0.41\\
	3&0.5&0.2&0.19&0.23&0.52&0.43&0.45&0.42&0.39&0.4&0.22&0.21\\
	3&0.55&0.16&0.31&0.53&0.5&0.53&0.44&0.44&0.36&0.2&0.19&0.36\\
	3&0.6&0.5&0.56&0.49&0.53&0.46&0.43&0.45&0.44&0.4&0.38&0.35\\
	3&0.65&0.23&0.53&0.58&0.5&0.52&0.45&0.53&0.41&0.41&0.21&0.36\\
	3&0.7&0.52&0.61&0.63&0.52&0.44&0.45&0.43&0.42&0.38&0.21&0.19\\
	3&0.75&0.45&0.47&0.66&0.56&0.42&0.45&0.4&0.41&0.26&0.37&0.18\\
	3&0.8&0&0.61&0.6&0.43&0.51&0.32&0.43&0.22&0.19&0.2&0.19\\
	3&0.85&0.47&0.72&0.57&0.45&0.45&0.35&0.51&0.29&0.22&0.18&0.19\\
	3&0.9&0.77&0.62&0.59&0.46&0.38&0.55&0.37&0.29&0.23&0.19&0.28\\
	3&0.95&0.77&0.66&0.42&0.42&0.48&0.48&0.52&0.39&0.39&0.23&0.49\\ \hline
	4&0&0.55&0.66&0.21&0.22&0.11&0.21&0.22&0.33&0.28&0.21&0.2\\
	4&0.05&0.48&0.44&0.45&0.33&0.22&0.24&0.29&0.31&0.3&0.24&0.28\\
	4&0.1&0.34&0.24&0.16&0.09&0.21&0.23&0.26&0.22&0.26&0.25&0.31\\
	4&0.15&0.27&0.42&0.47&0&0.16&0.18&0.21&0.19&0.29&0.27&0.27\\
	4&0.2&0.21&0.49&0.15&0.31&0.24&0.23&0.35&0.24&0.26&0.25&0.29\\
	4&0.25&0.44&0.36&0.49&0.32&0.33&0.3&0.39&0.37&0.39&0.3&0.32\\
	4&0.3&0.29&0.21&0.28&0.46&0.34&0.4&0.34&0.39&0.41&0.17&0.31\\
	4&0.35&0.41&0.21&0.24&0.46&0.46&0.45&0.42&0.32&0.32&0.3&0.3\\
	4&0.4&0.56&0.41&0.24&0.24&0.33&0.44&0.44&0.34&0.36&0.17&0.32\\
	4&0.45&0.42&0.29&0.37&0.26&0.26&0.48&0.32&0.34&0.29&0.22&0.31\\
	4&0.5&0.42&0.63&0.51&0.28&0.44&0.39&0.43&0.33&0.36&0.39&0.25\\
	4&0.55&0.53&0.39&0.31&0.54&0.52&0.4&0.2&0.38&0.36&0.34&0.18\\
	4&0.6&0.36&0.36&0.46&0.59&0.44&0.33&0.34&0.34&0.18&0.18&0.31\\
	4&0.65&0.47&0&0.46&0.58&0.35&0.34&0.32&0.2&0.33&0.22&0.32\\
	4&0.7&0.47&0&0.41&0.47&0.21&0.33&0.3&0.19&0.24&0.19&0.3\\
	4&0.75&0.41&0.4&0.44&0.41&0.34&0.22&0.21&0.37&0.29&0.33&0.35\\
	4&0.8&0.38&0.47&0.58&0.42&0.38&0.33&0.31&0.36&0.2&0.41&0.33\\
	4&0.85&0.37&0.75&0.38&0.42&0.35&0.29&0.3&0.35&0.36&0.35&0.34\\
	4&0.9&0.82&0.44&0.35&0.43&0.33&0.36&0.2&0.18&0.29&0.36&0.34\\
	4&0.95&0.59&0.37&0.44&0.33&0.35&0.19&0.19&0.44&0.19&0.3&0.31 \\ \hline \hline
	\end{longtable}
}

{\scriptsize
	\begin{longtable}{c|c|ccccccccccc}\caption{Composition of silhouette scores during the grid-search. Here $\omega_1 = 1000$, $\omega_2 = 1$ and $\omega_6 = \omega_3$. The presented scores correspond to the average score between all clusters, before the silhouette clean-up. }\label{tab:c3} \\  \hline 
		\multirow{2}{*}{$\omega_3$}&\multirow{2}{*}{$\omega_4$}&\multicolumn{11}{c}{$1 - \omega_5$} \\
		&&0.2&0.25&0.35&0.45&0.55&0.65&0.75&0.8&0.85&0.9&0.95\\
		\hline
		\endfirsthead
		
		\caption{\textit{(Continued)} Composition of silhouette scores during the grid-search. Here $\omega_1 = 1000$, $\omega_2 = 1$ and $\omega_6 = \omega_3$. The presented scores correspond to the average score between all clusters, before the silhouette clean-up. } \\  \hline 
		\multirow{2}{*}{$\omega_3$}&\multirow{2}{*}{$\omega_4$}&\multicolumn{11}{c}{$1 - \omega_5$} \\
		&&0.2&0.25&0.35&0.45&0.55&0.65&0.75&0.8&0.85&0.9&0.95\\
		\hline
		\endhead
2&0&0.3&0.25&0.27&0.3&0.26&0.32&0.45&0.46&0.46&0.48&0.49\\
2&0.05&0.34&0.33&0.26&0.27&0.39&0.33&0.45&0.35&0.46&0.43&0.45\\
2&0.1&0.33&0.25&0.28&0.34&0.31&0.33&0.39&0.41&0.44&0.48&0.5\\
2&0.15&0.27&0.26&0.31&0.23&0.34&0.35&0.38&0.38&0.41&0.48&0.52\\
2&0.2&0.28&0.28&0.22&0.21&0.36&0.35&0.37&0.38&0.48&0.46&0.5\\
2&0.25&0.3&0.27&0.23&0.25&0.29&0.36&0.41&0.43&0.48&0.48&0.51\\
2&0.3&0.33&0.23&0.24&0.17&0.31&0.31&0.45&0.41&0.41&0.48&0.49\\
2&0.35&0.25&0.25&0.2&0.26&0.37&0.41&0.36&0.45&0.46&0.46&0.5\\
2&0.4&0.25&0.22&0.15&0.28&0.34&0.38&0.4&0.39&0.43&0.51&0.51\\
2&0.45&0.24&0.29&0.21&0.31&0.42&0.39&0.44&0.45&0.48&0.47&0.51\\
2&0.5&0.3&0.41&0.51&0.42&0.42&0.47&0.38&0.4&0.49&0.51&0.51\\
2&0.55&0.64&0.71&0.57&0.44&0.63&0.56&0.54&0.58&0.61&0.62&0.57\\
2&0.6&0.37&0.4&0.58&0.77&0.74&0.73&0.72&0.72&0.68&0.64&0.58\\
2&0.65&0.52&0.5&0.3&0.8&0.8&0.79&0.75&0.72&0.67&0.61&0.59\\
2&0.7&0.54&0.48&0.56&0.85&0.85&0.81&0.76&0.7&0.66&0.61&0.5\\
2&0.75&0.56&0.43&0.87&0.87&0.86&0.82&0.75&0.71&0.65&0.57&0.57\\
2&0.8&0.57&0.55&0.89&0.9&0.86&0.81&0.73&0.7&0.65&0.63&0.56\\
2&0.85&0.63&0.47&0.93&0.91&0.84&0.78&0.71&0.69&0.63&0.61&0.58\\
2&0.9&0.59&0.51&0.94&0.87&0.82&0.77&0.61&0.64&0.62&0.59&0.56\\
2&0.95&0.97&0.97&0.94&0.84&0.76&0.7&0.65&0.61&0.62&0.6&0.63\\ \hline
3&0&0.6&0.55&0.35&0.35&0.3&0.26&0.27&0.26&0.32&0.34&0.32\\
3&0.05&0.48&0.47&0.44&0.36&0.33&0.29&0.23&0.33&0.24&0.33&0.32\\
3&0.1&0.43&0.48&0.45&0.4&0.32&0.27&0.26&0.35&0.32&0.31&0.31\\
3&0.15&0.32&0.32&0.38&0.38&0.31&0.28&0.34&0.34&0.33&0.3&0.32\\
3&0.2&0.39&0.4&0.39&0.36&0.3&0.25&0.35&0.32&0.32&0.31&0.34\\
3&0.25&0.46&0.53&0.47&0.38&0.3&0.27&0.27&0.34&0.24&0.29&0.33\\
3&0.3&0.56&0.55&0.5&0.39&0.31&0.29&0.29&0.32&0.3&0.32&0.31\\
3&0.35&0.61&0.55&0.48&0.29&0.34&0.31&0.37&0.37&0.36&0.35&0.31\\
3&0.4&0.57&0.54&0.37&0.41&0.45&0.44&0.42&0.4&0.39&0.35&0.33\\
3&0.45&0.58&0.53&0.5&0.47&0.5&0.48&0.43&0.4&0.39&0.35&0.36\\
3&0.5&0.61&0.54&0.45&0.51&0.52&0.5&0.44&0.41&0.42&0.4&0.36\\
3&0.55&0.55&0.55&0.49&0.55&0.54&0.49&0.46&0.46&0.36&0.38&0.35\\
3&0.6&0.59&0.55&0.55&0.55&0.56&0.47&0.46&0.44&0.44&0.39&0.36\\
3&0.65&0.6&0.54&0.64&0.62&0.56&0.44&0.51&0.46&0.45&0.38&0.36\\
3&0.7&0.51&0.53&0.64&0.61&0.52&0.46&0.48&0.44&0.44&0.37&0.32\\
3&0.75&0.38&0.48&0.66&0.6&0.5&0.47&0.47&0.45&0.39&0.37&0.33\\
3&0.8&0.67&0.65&0.69&0.53&0.57&0.49&0.44&0.43&0.39&0.36&0.31\\
3&0.85&0.5&0.75&0.67&0.54&0.58&0.5&0.42&0.41&0.4&0.36&0.33\\
3&0.9&0.8&0.74&0.65&0.59&0.5&0.46&0.39&0.41&0.4&0.36&0.36\\
3&0.95&0.78&0.67&0.58&0.54&0.57&0.49&0.42&0.29&0.47&0.41&0.41\\ \hline
4&0&0.64&0.56&0.35&0.37&0.28&0.33&0.26&0.31&0.3&0.29&0.29\\
4&0.05&0.52&0.58&0.43&0.43&0.31&0.31&0.33&0.33&0.32&0.3&0.29\\
4&0.1&0.56&0.53&0.41&0.38&0.29&0.32&0.33&0.31&0.3&0.3&0.31\\
4&0.15&0.55&0.56&0.45&0.38&0.36&0.29&0.33&0.29&0.35&0.32&0.32\\
4&0.2&0.39&0.56&0.41&0.32&0.37&0.25&0.31&0.35&0.33&0.33&0.32\\
4&0.25&0.42&0.36&0.33&0.33&0.33&0.31&0.34&0.34&0.35&0.32&0.31\\
4&0.3&0.43&0.4&0.37&0.38&0.48&0.38&0.37&0.37&0.31&0.32&0.3\\
4&0.35&0.45&0.4&0.37&0.48&0.47&0.48&0.43&0.38&0.36&0.34&0.31\\
4&0.4&0.43&0.4&0.47&0.46&0.48&0.49&0.42&0.41&0.39&0.34&0.33\\
4&0.45&0.43&0.41&0.43&0.54&0.5&0.53&0.43&0.41&0.38&0.34&0.33\\
4&0.5&0.51&0.58&0.47&0.51&0.48&0.5&0.42&0.41&0.4&0.36&0.33\\
4&0.55&0.53&0.4&0.43&0.52&0.57&0.46&0.42&0.42&0.4&0.36&0.32\\
4&0.6&0.38&0.47&0.58&0.64&0.55&0.48&0.45&0.41&0.37&0.36&0.32\\
4&0.65&0.53&0.55&0.54&0.59&0.54&0.48&0.46&0.41&0.38&0.34&0.33\\
4&0.7&0.52&0.5&0.54&0.62&0.54&0.49&0.44&0.39&0.36&0.33&0.33\\
4&0.75&0.57&0.56&0.57&0.61&0.52&0.46&0.42&0.41&0.37&0.35&0.37\\
4&0.8&0.59&0.59&0.7&0.6&0.54&0.46&0.43&0.43&0.37&0.35&0.32\\
4&0.85&0.53&0.81&0.64&0.59&0.5&0.45&0.41&0.38&0.4&0.35&0.33\\
4&0.9&0.86&0.74&0.62&0.55&0.47&0.44&0.39&0.38&0.36&0.37&0.36\\
4&0.95&0.79&0.68&0.6&0.5&0.47&0.39&0.42&0.42&0.41&0.41&0.44 \\ \hline \hline
	\end{longtable}
}

{\scriptsize
	\begin{longtable}{c|c|ccccccccccc}\caption{Composition of silhouette scores during the grid-search. Here $\omega_1 = 1000$, $\omega_2 = 1$ and $\omega_6 = \omega_3$. The presented scores correspond to the minimum score between all clusters, before the silhouette clean-up. }\label{tab:c4} \\  \hline 
		\multirow{2}{*}{$\omega_3$}&\multirow{2}{*}{$\omega_4$}&\multicolumn{11}{c}{$1 - \omega_5$} \\
		&&0.2&0.25&0.35&0.45&0.55&0.65&0.75&0.8&0.85&0.9&0.95\\
		\hline
		\endfirsthead
		
		\caption{\textit{(Continued)} Composition of silhouette scores during the grid-search. Here $\omega_1 = 1000$, $\omega_2 = 1$ and $\omega_6 = \omega_3$. The presented scores correspond to the minimum score between all clusters, before the silhouette clean-up.  } \\  \hline 
		\multirow{2}{*}{$\omega_3$}&\multirow{2}{*}{$\omega_4$}&\multicolumn{11}{c}{$1 - \omega_5$} \\
		&&0.2&0.25&0.35&0.45&0.55&0.65&0.75&0.8&0.85&0.9&0.95\\
		\hline
		\endhead
2&0&0.28&0.13&0.04&-0.03&-0.01&0&0.2&0.42&0.29&0.46&0.49\\
2&0.05&0.3&0.33&0.16&0.02&0.12&-0.04&0.21&0.07&0.43&0.19&0.27\\
2&0.1&0.26&0.13&0.15&0.21&0.02&-0.04&0.1&0.13&0.42&0.46&0.48\\
2&0.15&0.15&0.15&0.25&0.03&0.13&0.05&0.1&0.11&0.15&0.47&0.5\\
2&0.2&0.15&0.17&0.13&0.03&0.21&0.06&0.1&0.1&0.39&0.24&0.49\\
2&0.25&0.27&0.16&0.15&0.18&0.11&0.11&0.11&0.15&0.38&0.47&0.49\\
2&0.3&0.25&0.16&0.15&0.03&0.16&0.03&0.22&0.14&0.15&0.45&0.45\\
2&0.35&0.17&0.15&0.16&0.11&0.15&0.06&0.05&0.19&0.39&0.44&0.32\\
2&0.4&0.17&0.17&-0.04&0&-0.04&0.02&0.08&0.13&0.18&0.5&0.4\\
2&0.45&0.16&0.14&0.01&-0.02&-0.01&0.05&0.15&0.42&0.47&0.45&0.49\\
2&0.5&0.05&0.03&0.04&-0.06&-0.02&0.17&0.07&0.13&0.46&0.49&0.49\\
2&0.55&0.14&0.23&0.04&-0.06&0.62&0.51&0.5&0.58&0.53&0.61&0.57\\
2&0.6&0.07&0.11&0.02&0.76&0.72&0.68&0.7&0.68&0.63&0.63&0.58\\
2&0.65&0.31&0.25&0.19&0.8&0.78&0.77&0.71&0.68&0.58&0.61&0.58\\
2&0.7&0.48&0.27&0.07&0.84&0.85&0.79&0.71&0.67&0.64&0.61&0.31\\
2&0.75&0.55&0.41&0&0.87&0.85&0.78&0.71&0.66&0.58&0.44&0.56\\
2&0.8&0.34&0.36&0.89&0.89&0.85&0.79&0.68&0.66&0.65&0.62&0.51\\
2&0.85&0.47&0.32&0.92&0.91&0.8&0.74&0.65&0.65&0.6&0.53&0.54\\
2&0.9&0.33&0.45&0.94&0.86&0.78&0.71&0.46&0.64&0.49&0.49&0.53\\
2&0.95&0.97&0.96&0.93&0.77&0.69&0.62&0.65&0.6&0.62&0.52&0.63\\ \hline
3&0&0.45&0.48&-0.26&0.11&-0.04&0&0.04&0.11&0.18&0.21&0.17\\
3&0.05&0.36&0.44&0.03&0.25&-0.03&0.05&0&0.11&0.05&0.19&0.2\\
3&0.1&0.34&0.45&0.24&0.09&-0.01&0.01&0.03&0.14&0.17&0.12&0.14\\
3&0.15&-0.07&0.09&0.06&0.12&-0.01&0.04&0.09&0.07&0.11&0.12&0.19\\
3&0.2&0.18&0.11&0&0.09&-0.02&-0.02&0.07&0.17&0.12&0.15&0.24\\
3&0.25&0.05&0.07&0.18&0.11&0&0.01&0.09&0.13&0.05&-0.05&0.21\\
3&0.3&0.2&0.03&-0.01&-0.02&-0.04&0.11&0.14&0.22&0.25&-0.04&-0.02\\
3&0.35&-0.02&-0.05&-0.04&-0.17&0.24&0.01&0.34&0.21&0.15&0.3&0.02\\
3&0.4&-0.13&-0.05&0.05&0.36&0.34&0.3&0.24&0.3&0.33&0.25&0.29\\
3&0.45&-0.1&-0.08&0.39&0.38&0.39&0.36&0.37&0.22&0.25&0.34&0.02\\
3&0.5&0&-0.03&-0.01&0.42&0.23&0.37&0.37&0.31&0.28&0.09&0.05\\
3&0.55&-0.08&0&0.43&0.48&0.45&0.34&0.4&0.33&-0.05&0.08&0.23\\
3&0.6&-0.06&0.5&0.44&0.46&0.38&0.38&0.44&0.4&0.35&0.22&0.26\\
3&0.65&-0.06&0.52&0.56&0.44&0.48&0.39&0.49&0.41&0.39&0.03&0.19\\
3&0.7&0.4&0.39&0.56&0.45&0.43&0.42&0.42&0.29&0.33&0.05&-0.01\\
3&0.75&0.38&0.47&0.55&0.49&0.42&0.28&0.34&0.34&0.06&0.28&0\\
3&0.8&0&0.57&0.6&0.43&0.51&0.08&0.31&0.07&0.08&0.02&-0.01\\
3&0.85&0.32&0.71&0.57&0.45&0.45&0.06&0.02&0.05&0.05&0.09&-0.03\\
3&0.9&0.77&0.62&0.59&0.46&0.36&0.06&0&0.05&0.08&0&0.01\\
3&0.95&0.33&0.29&0.42&0.42&0.48&0.29&0.07&-0.1&0.39&0.03&0.02\\ \hline
4&0&0.55&-0.06&0.16&0.04&-0.11&0.05&0&0.17&0.12&0.05&0.04\\
4&0.05&-0.12&-0.01&0.03&0.12&0.02&0.04&0.16&0.12&0.09&0.06&0.12\\
4&0.1&0.07&-0.03&0.11&-0.11&-0.08&0.07&0.08&0.07&0.19&0&0.13\\
4&0.15&-0.06&-0.03&0.06&-0.12&0.03&-0.01&0.06&0.01&0.23&0.08&0.11\\
4&0.2&0.05&-0.06&0.16&-0.16&0.03&-0.13&0.09&0.05&0.18&0.15&0.12\\
4&0.25&-0.17&-0.12&-0.19&-0.14&-0.06&-0.14&-0.14&0.15&0.07&0.17&0.2\\
4&0.3&0.07&-0.01&0.21&-0.07&0.25&-0.05&-0.11&0.29&0.01&0.03&0.13\\
4&0.35&-0.11&-0.04&0.06&0.35&0.36&0.36&0.32&0.21&0.23&0.19&0.14\\
4&0.4&-0.08&0.01&0.08&0.17&0.29&0.33&0.3&0.28&0.27&0.07&0.18\\
4&0.45&-0.08&-0.12&0.2&0.18&0.14&0.42&0.25&0.19&0.18&0.08&0.16\\
4&0.5&0.24&0.51&0.21&-0.04&0.11&0.3&0.34&0.29&0.29&0.24&0.07\\
4&0.55&0.49&-0.07&0.1&0.41&0.4&0.29&0.13&0.31&0.25&0.23&0.04\\
4&0.6&0.16&0.14&0.37&0.57&0.34&0.21&0.26&0.27&0.11&0.13&0.17\\
4&0.65&0.38&0&0.35&0.31&0.28&0.2&0.24&0.11&0.23&0.03&0.16\\
4&0.7&0.3&0&0.25&0.44&0.19&0.24&0.18&0.06&0.06&0.05&0.16\\
4&0.75&0.31&0.34&0.12&0.36&0.28&0.13&0.13&0.26&0.19&0.19&0.2\\
4&0.8&0.27&0.31&0.58&0.41&0.3&0.26&0.21&0.34&0.06&0.09&0.16\\
4&0.85&0.04&0.75&0.38&0.41&0.32&0.2&0.22&0.11&0.28&0.1&0.17\\
4&0.9&0.82&0.31&0.35&0.39&0.23&0.28&0.1&0.14&0.17&0.16&0.12\\
4&0.95&0.59&0.31&0.33&0.27&0.29&0.06&0.09&0.3&0.1&0.18&0.23\\ \hline \hline
	\end{longtable}
}

\end{appendices}

\end{document}